\documentclass[11pt]{article}
\pdfoutput=1

\usepackage[utf8]{inputenc}
\usepackage{multirow}
\usepackage{amsmath, amsfonts, amssymb}
\usepackage{comment}
\usepackage{graphicx}
\usepackage{float}
\usepackage{psfrag}
\usepackage{amsthm}
\usepackage{caption}
\usepackage{subcaption}
\usepackage{import}
\usepackage{bm}
\usepackage[usenames,dvipsnames,svgnames,table]{xcolor}
\usepackage{enumerate}
\usepackage{arydshln}
\usepackage{soul}
 \usepackage{slashed}
\usepackage{mathrsfs}
\usepackage{mathalfa}
\usepackage{longtable}
\allowdisplaybreaks
 \usepackage{a4wide}
  \usepackage{tikz}
  \usepackage{tikz-cd}
  \usetikzlibrary{shapes.geometric,arrows}
  \usepackage{tcolorbox}
\tikzstyle{startstop} = [rectangle, rounded corners, minimum width=3cm, minimum height=1cm,text centered, draw=black, fill=red!30]
\tikzstyle{io} = [trapezium, trapezium left angle=70, trapezium right angle=110, minimum width=3cm, minimum height=1cm, text centered, draw=black, fill=blue!30]
\tikzstyle{process} = [rectangle, minimum width=3cm, minimum height=1cm, text centered, draw=black, fill=orange!30]
\tikzstyle{decision} = [diamond, minimum width=3cm, minimum height=1cm, text centered, draw=black, fill=green!30]
\tikzstyle{arrow} = [thick,->,>=stealth]
\tikzcdset{row sep/normal=0.75cm}

\usepackage[labelformat=simple]{subcaption}    %%Adding option to remove parenthesis
  \usepackage{color}
  \definecolor{dark-gray}{gray}{0.20}
  \definecolor{gray}{gray}{0.30}
  \definecolor{light-gray}{gray}{0.80}
  \definecolor{dark-red}{rgb}{0.7,0,0}
  \definecolor{dark-green}{rgb}{0,0.5,0}
  \definecolor{dark-blue}{rgb}{0.3,0.3,0.7}
  \definecolor{light-blue}{rgb}{0.8,0.8,1}
      \definecolor{swamp}{RGB}{240, 199, 197}

  \usepackage{pifont}

\usepackage{newunicodechar} % To write the definition on the next line
\usepackage{setspace}

\newcommand{\be}{\begin{equation}}
\newcommand{\ee}{\end{equation}}
\newcommand{\eq}[1]{(\ref{#1})}
\def\be{\begin{equation}}
\def\ee{\end{equation}}
\def\bea{\begin{eqnarray}}
\def\eea{\end{eqnarray}}

\newcommand{\beq}{\begin{equation}}  \newcommand{\eeq}{\end{equation}}
\newcommand{\bal}{\begin{aligned}}   \newcommand{\eal}{\end{aligned}}
\def\beqa{\begin{eqnarray}}
\def\eeqa{\end{eqnarray}}

\DeclareMathOperator{\diff}{\mathrm{d}}
\usepackage{stackengine}
\usepackage{calc}
\newlength\shlength
\newcommand\vv[2][0]{\setlength\shlength{#1pt}%
  \stackengine{-5.6pt}{$#2$}{\smash{$\kern\shlength%
    \stackengine{7.55pt}{$\mathchar"017E$}%
      {\rule{\widthof{$#2$}}{.57pt}\kern.4pt}{O}{r}{F}{F}{L}\kern-\shlength$}}%
      {O}{c}{F}{T}{S}}

\def\simleq{\; \raise0.3ex\hbox{$<$\kern-0.75em
      \raise-1.1ex\hbox{$\sim$}}\; }
   \def\simgeq{\; \raise0.3ex\hbox{$>$\kern-0.75em
      \raise-1.1ex\hbox{$\sim$}}\; }

\numberwithin{equation}{section}

\usepackage{jheppub}
\usepackage{hyperref}
\usepackage{cleveref}

\hypersetup{
	colorlinks=true,
	linkcolor=dark-blue,
	citecolor=dark-red,
	urlcolor=dark-green,
	linktoc=page
}

\theoremstyle{remark}

\crefname{appendix}{Appendix}{Appendices}

\title{\centering Cosmological Chameleons, String Theory and the Swampland}

\author{Gonzalo F. Casas$^1$,} \author{Miguel Montero$^{1}$,} \author{Ignacio Ruiz$^{1,\,2}$} 

\affiliation{$^1$Instituto de F\'{i}sica Te\'{o}rica UAM-CSIC, Universidad Aut\'{o}noma de Madrid, Cantoblanco, 28049 Madrid, Spain}
\affiliation{$^2$Departamento de F\'{i}sica Te\'{o}rica, Universidad Aut\'{o}noma de Madrid, Cantoblanco, 28049 Madrid, Spain}

\preprint{IFT-UAM/CSIC-24-71}
\emailAdd{gonzalo.f.casas@csic.es}
\emailAdd{miguel.montero@csic.es}
\emailAdd{ignacio.ruiz@uam.es}

\abstract{We study a scenario with a transient phase of cosmological acceleration that could potentially be realized in asymptotic corners of String Theory moduli space. A very steep scalar potential is temporarily stabilized by the effect of a nonzero density of heavy states, leading to acceleration, in what amounts to a cosmological version of the Chameleon mechanism. The density of heavy states is diluted by cosmological expansion, weakening their effect. After roughly one $e$-fold their effect can no longer stabilize the potential, and the accelerating phase ends. We also study a scenario where there is no potential and the transient acceleration is achieved by the counterbalancing effects of light and heavy towers of states. In both cases we show that it is not possible to obtain more than $\mathcal{O}(1)$ $e$-folds without transplanckian field excursions.  We also discuss the general EFT constraints on these models and explore a number of first attempts at concrete embeddings of the scenario in String Theory. These all turn out to face significant challenges.
}

\begin{document}
\hypersetup{pageanchor=false}
\makeatletter
\let\old@fpheader\@fpheader

\makeatother
\maketitle

\newcommand{\remove}[1]{\textcolor{red}{\sout{#1}}}

\pagenumbering{roman}

\newpage
\pagenumbering{arabic}
\setcounter{page}{1} 
\section{Introduction}
\label{sec:intro}

Considerable effort has been invested in building cosmological models that are integrated into UV-complete quantum gravity theories. One clear working arena for this problem is String Theory (see \cite{Flauger:2022hie,Cicoli:2023opf} and references therein for an overview on the state of the art cosmological string model building in String theory). Perhaps the simplest viable model is a cosmological constant, i.e. a de Sitter (dS) vacuum. While several top-down constructions have been proposed to realize a dS vacuum from String Theory (see \cite{Flauger:2022hie,Cicoli:2023opf} for extensive reviews), their status is not fully clear as of now, and none of them have been shown yet to feature complete control (see \cite{Balasubramanian:2005zx,deAlwis:2014wia,Polchinski:2015bea,Sethi:2017phn,Gao:2020xqh,Bena:2020xrh,Lust:2022lfc,Grana:2022nyp,Blumenhagen:2022dbo,Lust:2022xoq} however for recent progress in the KKLT scenario \cite{Kachru:2003aw}). In the context of the Swampland Program \cite{Vafa:2005ui,Brennan:2017rbf,Palti:2019pca,vanBeest:2021lhn,Grana:2021zvf,Harlow:2022ich,Agmon:2022thq,VanRiet:2023pnx},  this difficulty in finding explicit dS vacua has led to the so-called dS Swampland conjecture \cite{Obied:2018sgi}, which posits the absence of such vacua in the asymptotic regime of moduli space.
 
In light of this state of affairs, it is advisable to look for alternatives to a cosmological constant.  There is even more reason to do so after the publication of new results by the DESI Collaboration \cite{DESI:2024mwx} which show a mild preference for a time-varying dark energy; in other words, the cosmological constant might not be actually ``constant'', but change over time. Perhaps the simplest way to model a time-varying dark energy is via quintessence, where the dynamics of rolling scalar field(s) can result in accelerated expansion if the potential is flat enough (see \cite{Tsujikawa:2013fta, Achucarro:2018vey} for reviews). 

From a bottom-up perspective, the simplest quintessence model is one where the scalar field rolls on forever or is otherwise extremely long-lived -- what we might call ``eternal quintessence''. The situation with eternal quintessence models in String Theory is similar to dS. As it will be explained in more detail in Section \ref{S.CosmoCh}, Swampland conjectures forbid eternal quintessence solutions, since these eventually get to the asymptotic region of moduli space, where the potential is too steep to allow for acceleration.

On the other hand, there are no such constraints on \emph{transient acceleration} quintessence models, where the accelerated phase just lasts for a finite amount of time. Furthermore, the current phase of dark energy domination has only lasted for around $N\sim 0.5$ $e$-folds, so to explain current observations via a quintessence model we do not need the accelerating phase to last for very long. This contrasts with the case of inflation, where the number of $e$-folds to be supported is $N_{\rm inf}\sim\mathcal{O}(10)-\mathcal{O}(100)$ \cite{Planck:2018vyg}. 

Recently, reference \cite{Gomes:2023dat} studied several bottom-up models of transient acceleration thoroughly, showing their phenomenological viability as well as their compliance with Swampland constraints.  The goal of this paper is to explore in some detail a couple of concrete avenues for the stringy embedding of such models, by combining a modulus field $\phi$ (with a very steep or no potential) together with a heavy field coming from the lightest, stable element of a heavy tower of states. Heavy states (in towers or not) are ubiquitous in String Theory. In some cases (for instance, think of a D3-brane on a Calabi-Yau very close to the conifold point), the lightest state of a tower can be made very light and is correspondingly described by a light field $\chi$ that is part of the EFT. The effective field theory of the modulus and light towers such as these may be enough to generate a phase of transient acceleration as in \cite{Gomes:2023dat}. When the field $\chi$ has a mass $m_\chi(\phi)$ that grows with $\phi$, the mass term $m_\chi^2\chi^2$ can balance a steep $\phi$-dependent potential, creating a local minimum for a while. While this minimum exists, the cosmology is accelerated. However, since the vev of the scalar $\chi$ dilutes with time due to accelerated expansion, eventually the minimum disappears, and the accelerated epoch ends.

In a sense, the idea that we will study here is a cosmological version of the Chameleon mechanism \cite{Khoury:2003rn,Khoury:2003aq}, where a runaway scalar is stabilized locally via its coupling to matter in regions of high matter density. In our setup, the role of the ``matter density'' is played by the field $\chi$, which permeates the whole Universe. Because of this, we will refer to the scenario we consider as ``Cosmological Chameleon''.

In this paper we only carry the first steps to explore this setup: We will study general bottom-up constraints and limitations of the scenario (for instance, we will find that the accelerated phase can never last more than $\mathcal{O}(1)$ $e$-folds), quantify the limitations of the effective field theory we use, and describe our attempts to find a concrete stringy embedding of the scenario. While we failed to do this, we describe in each case the obstacles that would have to be overcome to achieve a successful embedding.

The structure of the paper is as follows. In section \ref{S.CosmoCh}, we present the model we will study and discuss the rationale for its top-down origin, as well as its compatibility with Swampland constraints (which was already pointed out in \cite{Gomes:2023dat}). In addition, we explore the cosmological viability and derive results from two setups: one involving a runaway potential coupled to a heavy species, and the other involving a light tower of states combined with the heavy states.
In section \ref{s: validity EFT} we examine the various caveats and limitations of the scenario. Section \ref{section 4: examplecitos} discusses our attempts at a stringy embedding, why they are incomplete, and the general difficulties encountered. Finally, in section \ref{s: conclusions} we encapsulate our findings and present forward-looking outlooks.

\section{Cosmological Chameleons}\label{S.CosmoCh}

 We will begin by describing our general setup. Consider a low-energy action in $d$ spacetime dimensions, consisting of the Einstein-Hilbert term plus some sets of (canonically normalized\footnote{While in general it is not possible to put all scalars in canonical form, for our purposes we will mostly work with a single scalar, which we can always normalize. For some comments on generalization to more scalars that need not be canonically normalized, see Section \ref{S.SeveralScalars}.}) scalars $\{\phi^i\}_{i=1}^K$ and $\{\chi_I\}_I$. The $\phi^i$ are the usual moduli fields encountered in string compactifications -- volumes of cycles and the string coupling--, while the $\chi_I$ are meant to be the new ingredient we consider in this paper -- heavier states, such as wrapped D-branes, etc -- which are not part of the usual field theory. We will also consider a potential $V(\vec\phi)$ for the moduli, obtained from the usual ingredients -- fluxes, quantum and higher derivative corrections, etc 

Importantly, the vacuum expectation values (vev) of the scalars $\vec{\phi}$ parameterize the masses of the scalars $\{\chi_I\}_I$; one typically obtains exponential dependencies in String Theory.  The action takes the form:\footnote{One could also include fermions that become heavy in some asymptotic limit, and that contribute to the scalar potential via a fermion condensate. We will not entertain that possibility in this paper.}
\begin{align}\label{l.action}
    S^{(d)}&\supseteq\frac{1}{2}\int{\rm d}^d x\sqrt{-g}\left\{\kappa_d^{-2}\left[\mathcal{R}_g-\sum_{i=1}^K(\partial\phi^i)^2\right]-2V(\vec\phi)-\sum_Im_I(\vec\phi)^2\chi_I^2-\sum_I(\partial\chi_I)^2\right\}\notag\\
    &=\frac{1}{2}\int{\rm d}^dx\sqrt{-g}\left\{\kappa_d^{-2}\left[\mathcal{R}_g-\sum_{i=1}^K(\partial\phi^i)^2\right]-2V_{\rm eff}(\vec\phi)-\sum_I(\partial\chi_I)^2\right\}\;,
\end{align}
where we introduce the effective potential
\begin{equation}\label{e:Veff}
    V_{\rm eff}(\vec\phi)=V(\vec\phi)+\frac{1}{2}\sum_Im_I(\vec\phi)^2\chi_I^2\;,
\end{equation}
From \eqref{l.action}, the stress energy tensor of a massive scalar $\chi_I$ is given by
\begin{equation}
    T^{(I)}_{\mu\nu}=\partial_\mu \chi_I\partial_\nu \chi_I-\frac{1}{2}g_{\mu\nu}\left[\left(\partial\chi_ I\right)^2+m_I(\vec\phi)^2\chi_I^2\right]\;.
\end{equation}
If one further assumes that $\chi_I$ is ``frozen'' (i.e. its kinetic term is negligible) and its mass scales as $m_I(\vec\phi)=m_{I,0} \exp \left\{\mu_1\phi^1+...+\mu_K\phi^K\right\}=m_{I,0}e^{\ \mu_i\phi^i}$, then we can identify the trace of the $T^{(I)}_{\mu\nu}$ with (minus) the scalar density, resulting in
\begin{equation}
    \rho_I\approx -T^{(I)\nu}_\nu(\vec\phi)=-T^{(I)\mu}_\mu(\vec\phi_0)\left[1+2\mu_i(\phi^i-\phi_0^i)\right]+\mathcal{O}(\vec\phi-\vec\phi_0)^2\;,
\end{equation}
thus allowing us to rewrite \eqref{e:Veff} as 
\begin{align}
    V_{\rm eff}(\vec\phi)&=V(\phi)-\frac 1d
\sum_I T^{(I)\nu}_\nu(\vec\phi)\notag\\
&=V(\phi)+\frac{1}{d}\sum_I\rho_{I,0}\left[1+2\mu_i(\phi^i-\phi_0^i)\right]+\mathcal{O}(\vec\phi-\vec\phi_0)^2\;,
\label{chcop}\end{align}
where we have assumed that the mass of all species $\{\chi_I\}$ scales in the same way (though this assumption will be relaxed in Section \ref{S.SeveralScalars}). The coupling \eqref{chcop}, where $\rho_{I,0}=\frac{1}{2}m_{I,0}^2\chi_I^2$, is precisely the key ingredient of the Chameleon mechanism \cite{Khoury:2003rn,Khoury:2003aq}; in that realization, $\rho_{0,I}$ is replaced e.g. by the local matter density in the Solar System, but the effect is similar; the scalar $\phi$ achieves a minimum. In the traditional Chameleon mechanism, this minimum ensures the absence of fifth-forces associated with the scalars $\vec{\phi}$; in our setup, the point of the minimum is to ensure an accelerated phase of the cosmology.  This is why we refer to the construction as a ``Cosmological Chameleon''. The difference with the astrophysical Chameleons relies on the fact that rather than the density corresponding to that of ordinary matter in denser clusters (e.g. our Solar System), here the massive scalars $\{\chi_I\}_I$ are homogeneous over space, evolving only with time.

In order to study the evolution of these fields, we take a cosmological ansatz consisting of a $d$-dimensional FLRW background, with metric
\begin{equation}
    {\rm d} s^2=-{\rm d} t^2+a(t)^2\left[\frac{{\rm d} r^2}{1-kr^2}+r^2{\rm d} \Omega_{d-2}^2\right]\;,
\end{equation}
 Hubble factor $H=\frac{\dot a}{a}=\partial_t\log a$, and spatially homogeneous fields. The equations of motion will then be given by\footnote{We take $k=0$, zero spatial curvature.}
\begin{subequations}\label{e.eom}
\begin{align}
    \Ddot{\phi}^i+(d-1)H\dot{\phi}^i+\kappa_d^{2}\partial^{\phi^i}V_{\rm eff}(\vec\phi,t)&=0,\tag{M$\phi^i$}\label{eq.phiEOM}\\
    \Ddot{\chi}_I+(d-1)H\dot{\chi}_I+m_I(\vec\phi)^2\chi_I&=0,\tag{M$\chi_I$}\\
    (d-1)(d-2)H^2-\left[\sum_{i=1}^K\dot{\phi}^i{}^2+\kappa_d^2\left(\sum_I\dot{\chi}_I^2+2V_{\rm eff}(\vec\phi,t)\right)\right]&=0,\tag{F1}\label{eq.F1}\\
    2(d-2)\dot H+(d-1)(d-2)H^2+\left[\sum_{i=1}^K\dot{\phi}^i{}^2+\kappa_d^2\left(\sum_I\dot{\chi}_I^2-2V_{\rm eff}(\vec\phi,t)\right)\right]&=0\tag{F2}.
\end{align}
\end{subequations}
The sum in $I$, which runs over several fields, all with identical $\phi$ dependence, is included to account for the fact that in general a whole tower of states, according to Swampland expectations \cite{Vafa:2005ui,Brennan:2017rbf,Palti:2019pca,vanBeest:2021lhn,Grana:2021zvf,Harlow:2022ich,Agmon:2022thq,VanRiet:2023pnx}, might have to be considered. More precisely, from the Swampland Distance Conjecture (SDC) \cite{Ooguri:2006in}, it is expected that along any (geodesic) trajectory  in moduli space $\mathcal{M}$ where the field $\vec{\phi}$ takes values, there exists a tower of states whose characteristic mass becomes light exponentially with the traveled distance as asymptotic regions of $\mathcal{M}$ are probed:
    \begin{equation}
        m(\Delta\phi)\sim m(0) e^{-\alpha\Delta\phi}\;,\quad \text{as }\;\Delta\phi\to\infty\;,\quad\text{with }\alpha=\mathcal{O}(1)\;.
    \end{equation}
This has been extensively tested in different string compactifications \cite{ Baume:2016psm,Klaewer:2016kiy, Blumenhagen:2017cxt, Grimm:2018ohb,Heidenreich:2018kpg, Blumenhagen:2018nts, Grimm:2018cpv, Buratti:2018xjt, Corvilain:2018lgw, Joshi:2019nzi,  Erkinger:2019umg, Marchesano:2019ifh, Font:2019cxq,  Gendler:2020dfp, Lanza:2020qmt, Klaewer:2020lfg, Rudelius:2023mjy}, as well as motivated from bottom-up arguments \cite{Calderon-Infante:2023ler,Ooguri:2024ofs,Aoufia:2024awo}. Connected to the SDC, the Emergent String Conjecture \cite{Lee:2019wij}, describes the nature of this tower of states: it is either a tower of perturbative string states, or a KK tower for a number of extra dimensions that are becoming large and that decompactify in the infinite distance limit.

The potential $V(\Vec{\phi})$ is also constrained by Swampland arguments. Precisely, the (asymptotic) de Sitter Swampland Conjecture \cite{Obied:2018sgi} constrains the gradient of the potential at asymptotic regions in moduli space
\begin{equation}\label{e.dS conj}
    \lambda =\frac{\|\nabla V(\Vec{\phi})\|}{V(\Vec{\phi)}}\geq c_d\;,\quad \text{as}\quad\Delta\phi\to\infty\;,
\end{equation}
where $c_d$ is some $\mathcal{O}(1)$ number depending only on the dimensionality $d$ of spacetime. The values of $c_d$ are in turn constrained by the strong dS conjecture, \cite{Rudelius:2021azq}, which requires $c_d^{\rm strong}=\frac{2}{\sqrt{d-2}}$. This value makes the asymptotic potential just steep enough to prevent accelerated expansion at late times, provided the equations of motion for $\vec\phi$ asymptotically follow a geodesic on $\mathcal{M}$, as is generally expected \cite{Calderon-Infante:2020dhm,Grimm:2022sbl,Calderon-Infante:2022nxb}. This conjecture has been extensively checked in asymptotic regions of moduli space \cite{Obied:2018sgi,Maldacena:2000mw,Hertzberg:2007wc,Andriot:2019wrs,Andriot:2020lea,Calderon-Infante:2022nxb,Shiu:2023fhb,Shiu:2023nph,Cremonini:2023suw,Hebecker:2023qke,VanRiet:2023cca,Seo:2024fki}. We will always consider exponential potentials with $\lambda\geq\frac{1}{d-2}$.

Finally, the Trans-Planckian Censorship Conjecture (TCC) \cite{Bedroya:2019snp} requires that accelerating vacua have a short lifetime,
\begin{equation}\label{e.TCC lifetime}
     t_{\rm accel.}\leq \frac{1}{H_0}\log\left(\frac{M_{\rm Pl,d}}{H_0}\right)\;,
\end{equation}
and that asymptotic runaway potentials have an exponential slope larger than $c_d^{\rm TCC}=\frac{2}{\sqrt{(d-1)(d-2)}}$. Note that in principle this would allow for accelerated expansion (just not \emph{too accelerated}).

In the following, we will study two kinds of Cosmological Chameleon scenarios near the boundary of the field space $\Vec{\phi}$. The first consists of two towers of states, a light tower and a heavy tower (in the sense of how their masses evolve as we move in some certain asymptotic direction in moduli space). The second case corresponds to a runaway scalar potential plus a heavy tower. Both will lead us to a frozen cosmological constant for about $N\lesssim1$ $e$-folds, consistent with the aforementioned swamplandish expectations. In both scenarios, we will also contrast with Swampland conjectures just described (and we will agree, whenever there is an overlap, with the conclusions of the more general analysis in \cite{Gomes:2023dat}). 

\subsection{Single scalar with light and heavy towers\label{s.light and heavy}}

We first consider the case in which we have a single moduli $\phi$ and two species $\psi$ and $\chi$, respectively becoming light and heavy with $\phi$ with exponential rates $\alpha$ and $\mu$, while $V(\phi)\equiv 0$. This way the effective potential is
\begin{equation}
    V_{\rm eff}(\phi)=\frac{1}{2}\left(M_0^2e^{-2\alpha\phi}\psi^2+m_0^2e^{2\mu\phi}\chi^2\right)\;,
\end{equation}
and the equations of motion are given by
\begin{subequations}
    \begin{align}
        \ddot\phi+(d-1)H\dot\phi+\kappa_d^2\left(\mu m_0^2e^{2\mu\phi}\chi^2-\alpha M_0^2e^{-2\alpha\phi}\psi^2\right)&=0\\
        \ddot\psi+(d-1)H\dot\psi+M_0^2e^{-2\alpha\phi}\psi&=0\\
         \ddot\chi+(d-1)H\dot\chi+m_0^2e^{2\mu\phi}\chi&=0\\
         H-\frac{1}{\sqrt{(d-1)(d-2)}}\sqrt{\dot\phi^2+\kappa_d^2\left(\dot\psi^2+\dot\chi^2+M_0^2e^{-2\alpha\phi}\psi^2+m_0^2e^{2\mu\phi}\chi^2\right)}&=0\;.
    \end{align}
\end{subequations}
We will assume that initially all the scalars are effectively frozen $\dot\phi\approx\dot\psi\approx\dot\chi\approx0$, with $\phi_0$ further stabilized at the minimum of $V_{\rm eff}(\phi)$, given by
\begin{equation}\label{e:phi0 light heave}
    \phi_0=\frac{1}{\alpha+\mu}\log\left(\sqrt{\frac{\alpha}{\mu}}\frac{M_0}{m_0}\left|\frac{\psi_0}{\chi_0}\right|\right)\;,
\end{equation}
so that the initial \emph{physical} masses of the light and heavy scalars are
\begin{equation}
    M_\psi=M_0\left(\sqrt{\frac{\alpha}{\mu}}\frac{M_0}{m_0}\left|\frac{\psi_0}{\chi_0}\right|\right)^{-\frac{\alpha}{\mu+\alpha}},\qquad m_\chi=m_0\left(\sqrt{\frac{\alpha}{\mu}}\frac{M_0}{m_0}\left|\frac{\psi_0}{\chi_0}\right|\right)^{\frac{\mu}{\mu+\alpha}}\;.
\end{equation}
Note that in this initial point the densities of both states are similar, as
\begin{equation}\label{e.initialEQ}
    \frac{\rho_{\psi,0}}{\rho_{\chi,0}}=\frac{M_\psi^2\psi_0^2}{m_\chi^2\chi_0^2}=\frac{\mu}{\alpha}=\mathcal{O}(1)\;,
\end{equation}
being exactly the same for $\alpha=\mu$.\footnote{One might naively think that dual towers always feature $\mu=\alpha$ for light and heavy states in any limit. This is not generally true when more than one modulus is involved such as e.g. a circle compactification where winding states have dependence on the dilaton, see \cite{Etheredge:2023odp}.} Using this we have
\begin{align}
    H_0&=\frac{\kappa_d}{\sqrt{(d-1)(d-2)}}\sqrt{M_\psi^2\psi_0^2+m_\chi^2\chi_0^2}=\frac{\kappa_d}{\sqrt{(d-1)(d-2)}}\sqrt{1+\frac{\mu}{\alpha}}m_{\chi}|\chi_0|,\\
    m_{\rm eff}^{(\phi)}& =\kappa_d^{-1}\sqrt{V_{\rm eff}''(\phi)|_{\phi_0}}=\sqrt{2\alpha\mu(d-1)(d-2)} H_0 \label{eq.mPhi1},
\end{align}
and the following equations of motion for this frozen, quasi-dS era : 
\begin{subequations}
    \begin{align}
       \phi-\phi_0&=\log\left(\sqrt{\frac{\alpha}{\mu}}\frac{\psi}{\chi}\frac{\chi_0}{\psi_0}\right)^{\frac{1}{\alpha+\mu}},\\
        \ddot\psi+M_\psi^2\left(\frac{\psi}{\chi}\frac{\chi_0}{\psi_0}\right)^{-\frac{2\alpha}{\mu+\alpha}}\psi&=0,\\
        \ddot\chi+m_\chi^2\left(\frac{\psi}{\chi}\frac{\chi_0}{\psi_0}\right)^{\frac{2\mu}{\mu+\alpha}}\chi&=0.
    \end{align}
\end{subequations}
The solution is
\begin{subequations}
    \begin{align}
    \psi(t)&\approx\psi_0(t)\left[1-\frac{M_\psi^2(t-t_0)^2}{2}\left|\frac{\psi_0}{\chi_0}\right|^{-\frac{2\alpha}{\alpha+\mu}}+\frac{m_\psi^4(t-t_0)^4}{24(\alpha+\mu)}\left|\frac{\psi_0}{\chi_0}\right|^{-\frac{4\alpha}{\alpha+\mu}}\left(\mu-\alpha+2\alpha\frac{m_\chi^2\psi_0^2}{M_\psi^2\chi_0^2}\right)\right],\\
        \chi(t)&\approx\chi_0(t)\left[1-\frac{m_\chi^2(t-t_0)^2}{2}\left|\frac{\psi_0}{\chi_0}\right|^{\frac{2\mu}{\alpha+\mu}}+\frac{m_\chi^4(t-t_0)^4}{24(\alpha+\mu)}\left|\frac{\psi_0}{\chi_0}\right|^{\frac{4\mu}{\alpha+\mu}}\left(\alpha-\mu+2\mu\frac{M_\psi^2\chi_0^2}{m_\chi^2\psi_0^2}\right)\right],\\
         \phi-\phi_0&\approx\frac{1}{2}\left(m_\chi^2\left(\frac{\psi_0}{\chi_0}\right)^{\frac{2\mu}{\alpha+\mu}}-M_\psi^2\left(\frac{\psi_0}{\chi_0}\right)^{-\frac{2\alpha}{\alpha+\mu}}\right)(t-t_0)^2\;,
    \end{align}
\end{subequations}
written respectively as a power expansion in 
\begin{equation*}
    \mathcal{O}\left(M_\psi^2(t-t_0)^2\left|\frac{\psi_0}{\chi_0}\right|^{-\frac{2\alpha}{\alpha+\mu}}\right),\quad \mathcal{O}\left(m_\chi^2(t-t_0)^2\left|\frac{\psi_0}{\chi_0}\right|^{\frac{2\mu}{\alpha+\mu}}\right)    
    \end{equation*}
    and their sum. At least initially, $\phi$ will grow (for $\frac{\mu}{\alpha}<\left(\psi_0/\chi_0\right)^4$), decrease ($\frac{\mu}{\alpha}>\left(\psi_0/\chi_0\right)^4$) or remain with the same value ($\frac{\mu}{\alpha}=\left(\psi_0/\chi_0\right)^4$) depending the relation between initial conditions and exponential factors, depending on whether
    \begin{equation}
        \frac{M_\psi}{m_\chi}=\sqrt{\frac{\mu}{\alpha}} \left|\frac{\psi_0}{\chi_0}\right|^{-1}\lesseqqgtr \left|\frac{\psi_0}{\chi_0}\right|\;.
    \end{equation}
    The lifetime of this quasi-dS phase is given by the time the above series expansions take to break-down, namely
    \begin{align}
        t_{\rm dS}&\approx\min\left\{\frac{\sqrt{2}}{M_\psi}\left|\frac{\psi_0}{\chi_0}\right|^{\frac{\alpha}{\alpha+\mu}},\frac{\sqrt{2}}{m_\chi}\left|\frac{\psi_0}{\chi_0}\right|^{-\frac{\mu}{\alpha+\mu}}\right\}\;.
    \end{align}
    We will now estimate an upper bound on the number $N$ of $e$-folds this mechanism can support. As $H$ is approximately constant during this phase  (then it proceeds to decrease), $N\approx H_0t_{\rm dS}$, leading to
    \begin{equation}
        N\leq \sqrt{\frac{2(\alpha+\mu)}{(d-1)(d-2)}}|\psi_0|^{-\frac{\mu}{\alpha+\mu}}|\chi_0|^{\frac{\alpha+2\mu}{\alpha+\mu}}\min\left\{\alpha^{-1/2},\mu^{-1/2}\right\}\;.
    \end{equation}
    Irrespective of bounds on $\alpha$ and $\mu$, we find that in general for subplanckian $|\psi_0|,\,|\chi_0|\lesssim \kappa_d^{-1}$, we have
    \begin{equation}
        N\leq\frac{2}{\sqrt{(d-1)(d-2)}}\;, \label{eq:efoldLH}
    \end{equation}
    saturated for $\alpha=\mu$ and $|\psi_0|=|\chi_0|=\kappa_d^{-1}$. This prevents the possibility of a long-lived de Sitter phase through this mechanism, see Figure \ref{f.comp}. 
    
    The numerical coefficient in \eqref{eq:efoldLH}  is at face value the same as the TCC value $c^{\rm TCC}_d$. As far as we can tell, this is simply a coincidence, but as we will see around \eqref{eq.TCC again} the TCC nonetheless holds for these kinds of solutions, as the quasi-dS epochs are short-lived enough. 

\begin{figure}[H]
\begin{center}
\begin{subfigure}[b]{0.432\textwidth}
\center
\includegraphics[width=\textwidth]{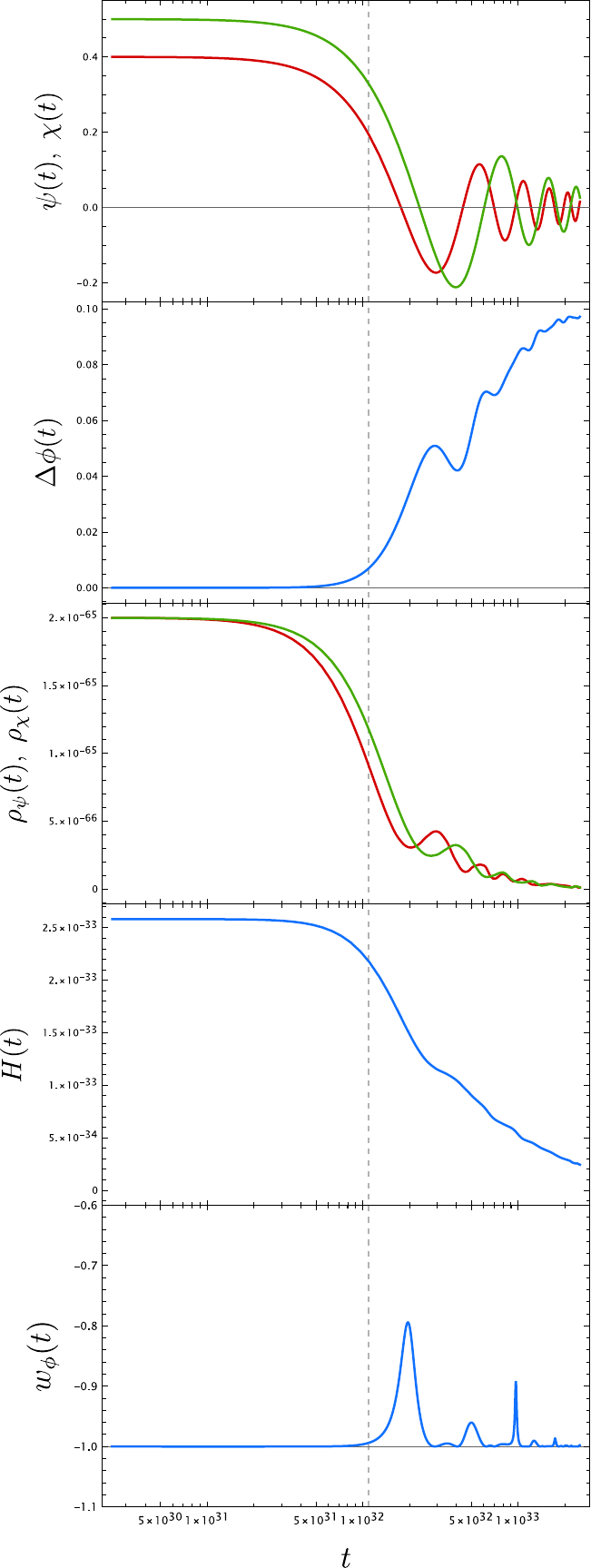}
\caption{Light and heavy states.} \label{f.PotLightPlots}
\end{subfigure}
\begin{subfigure}[b]{0.45\textwidth}
\center
\includegraphics[width=\textwidth]{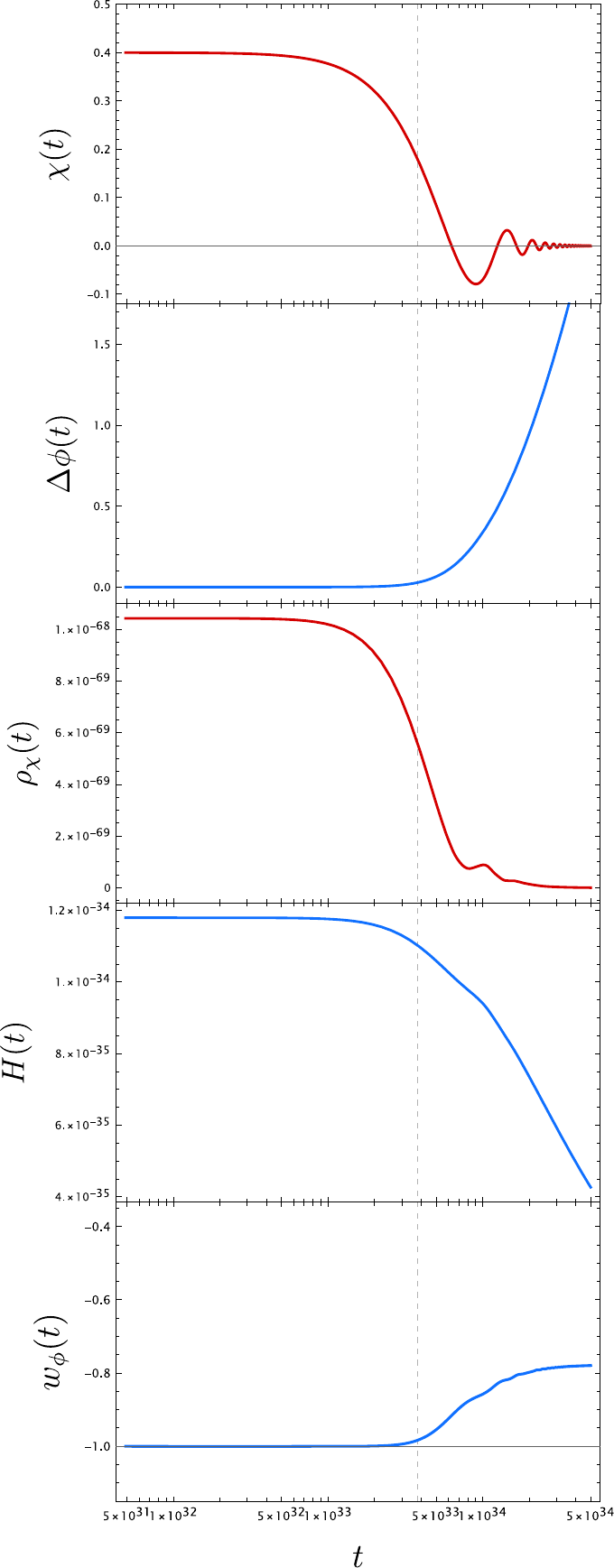}
\caption{Runaway potential and heavy species.} \label{f.PotHeavy}
\end{subfigure}
\caption{Massive field profiles, moduli distance traveled, massive field densities, Hubble parameter and equation of state of $\phi$ as a function of time for the light and heavy fields and runaway potential and heavy field cases. The parameters used are $d=4$, $\alpha=\mu=\sqrt{\frac{d-1}{d-2}}=\sqrt{\frac{3}{2}}$, $M_0=8\cdot 10^{-33}$, $m_0=1.25\cdot 10^{-32}$, $\psi_0=0.5$ and $\chi_0=0.4$ for light (green) and heavy (red) states (\ref{f.PotLightPlots}), and $d=4$, $\lambda=\frac{2}{\sqrt{(d-1)(d-2)}}=\sqrt{\frac{2}{3}}$, $\mu=\sqrt{\frac{d-1}{d-2}}=\sqrt{\frac{3}{2}}$, $m_0=3.61\cdot 10^{-34}$, $V_0=3.13\cdot10^{-68}$ and $\chi_0=0.4$ for a runaway potential with a heavy species (\ref{f.PotHeavy}). The estimated end of the quasi-dS phase is indicated with a dashed line and quantities are given in Planck units.
\label{f.plots}}
\end{center}
\end{figure}

    See Figures \ref{f.PotLightPlots} and \ref{f.LightHeavy3D} for numerical solutions of this scenario, which asymptote to $\phi\to \infty$ after the quasi-dS era. The last plot of Figure \ref{f.LightHeavy3D} depicts the equation of state of our scalar/dark energy
    \begin{equation}\label{e.wphi}
        w_\phi(t)=\frac{\frac{1}{2}\dot\phi^2-\kappa_d^2V_{\rm eff}(\phi,t)}{\frac{1}{2}\dot\phi^2+\kappa_d^2V_{\rm eff}(\phi,t)}\,.
    \end{equation}

Even after the quasi-dS era ends, the plot does not deviate much from $w_\phi\equiv -1$, i.e. an actual cosmological \emph{constant}. The reason is that even though the minimum of the effective potential quickly disappears as $\chi$ and $\psi$ dilute, it does so uniformly in both directions, so that the scalar does not pick up enough velocity to raise the value of $w_\phi$ significantly.

\begin{figure}[h]
\begin{center}
\begin{subfigure}[b]{0.49\textwidth}
\center
\includegraphics[width=\textwidth]{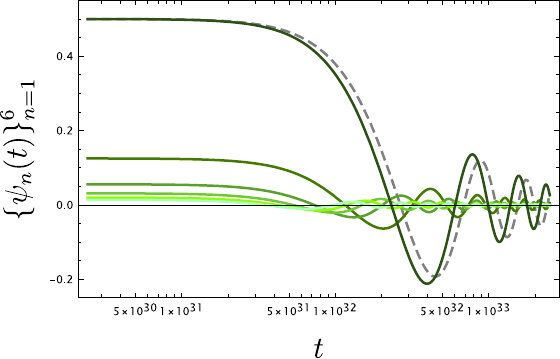}
\caption{Profile of the first six modes $\{\psi_n(t)\}_{n=1}^6$ of the light tower. The associated shade of green gets lighter as $n$ grows.} \label{f.PotLightPlotsMulti}
\end{subfigure}
\hfill
\begin{subfigure}[b]{0.49\textwidth}
\center
\includegraphics[width=\textwidth]{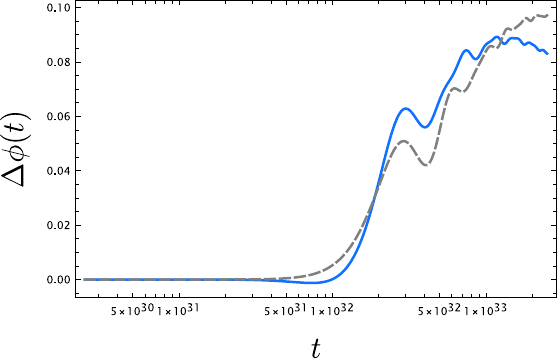}
\caption{Distance traveled by the canonically normalized scalar $\Delta\phi(t)$ considering both the six first states of the light tower or only the first one.} \label{f.plotsMultPhi}
\end{subfigure}
\begin{subfigure}[b]{0.49\textwidth}
\center
\includegraphics[width=\textwidth]{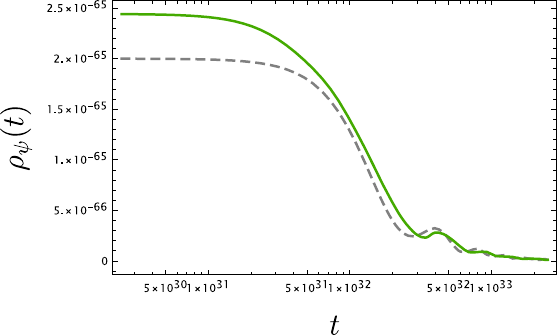}
\caption{Density of the light tower $\rho_{\psi}=\sum_i\rho_{\psi,i}$ considering the first six modes or only the first one.} \label{f.plotsMultDens}
\end{subfigure}
\caption{Profiles of the first six modes of the light tower, moduli distance traveled and total field densities of the light tower as a function of time for the light and heavy fields scenario. In gray dashed lines, the dynamics in the case of a single light field are depicted for comparison. The parameters used are the same as in Figure \ref{f.PotLightPlots}, with $\psi_{n,0}=\frac{\psi_0}{n^2}$. All quantities are given in Planck units.
\label{f.plotsMULTI}}
\end{center}
\end{figure}

For the above analysis we have only considered a single light and heavy state. The time a given light or heavy scalar remains ``frozen'' is inversely proportional to their mass, so the cosmological evolution will be dominated by the first state of the heavy and light towers, which anyway are the ones expected to have longest lifetime once intra-tower decays are considered. Even assuming full tower stability (see \cite{Etheredge:2023usk} for discussion on heavy towers), populating higher tower levels  just introduces small corrections to our results. For illustration, the total density of a light KK-like tower is 
    \begin{equation}
        \rho_\psi=\sum_n\rho_{\psi_n}\approx \sum_n\psi_n^2m_{\psi_n}^2=m_0^2\sum_n\psi_n^2n^2\;.
    \end{equation}
    All the above expressions remain the same, simply replacing 
    \begin{equation}
    |\psi_0|\to\sqrt{\sum_nn^2\psi_{n,0}^2}.
    \end{equation}
    As an example, for a $\psi_{n,0}\sim\frac{\psi_0}{n^2}$ dependence, taking into account the full tower of states amounts to a $\mathcal{O}(1)$ factor inside the $\log$ argument in \eqref{e:phi0 light heave} smaller than $\frac{\pi}{\sqrt{6}}\approx1.2826$. In Figure \ref{f.plotsMULTI} the light-heavy tower system is solved for both several and a single state in the light tower, with the quantitative behavior of the quasi-dS phase not changing much.

In summary,  we get an $\mathcal{O}(1)$ number of $e$-folds, assuming that $\vert\chi_0\vert$ and $|\psi_0|$ can be taken at most to be of order 1 in Planck units. A Planckian or transplanckian field range often results in higher order corrections to the effective action becoming relevant. We will discuss the limits of validity and approximation of this construction in Section \ref{s: validity EFT}.

\subsection{Single scalar with runaway potential and heavy tower\label{s.potential and heavy}}
The second case we will consider is that of a single scalar running down an exponential potential $V(\phi)=V_0e^{-\lambda\phi}$ and encountering a heavy (stable) scalar state $\chi$, with mass $m(\phi)=m_0e^{\mu\phi}$. Here the effective potential for $\phi$ is 
\begin{equation}
    V_{\rm eff}(\phi,t)=V_0e^{-\lambda\phi}+\frac{1}{2}m_0^2e^{2\mu\phi}\chi^2.\;\label{eq: runheavy}
\end{equation}
The equations of motion are then given by
\begin{subequations}\label{eom1}
\begin{align}
    \ddot{\phi}+(d-1)H\dot{\phi}-\kappa_d^{2}\left[\lambda V_0e^{-\lambda\phi}+\mu m_0^2e^{2\mu\phi}\chi(t)^2\right]&=0\label{eom1phi},\\
    \ddot{\chi}+(d-1)H\dot\chi+m_0^2e^{2\mu\phi}\chi&=0\label{eom1chi},\\
    (d-1)(d-2)H^2-\left[\dot\phi^2+\kappa_d^2\left(\dot\chi^2+2V_0e^{-\lambda\phi}+m_0^2e^{2\mu\phi}\chi^2\right)\right]&=0.
\end{align}
\end{subequations}
As in the previous case, we will consider initial conditions in which $\chi$ is frozen and $\phi_0$ is stabilized at the minimum of the effective potential, corresponding to  
\begin{equation}\label{eq: phi0}
  \phi^{(0)}(\chi)=\underbrace{\frac{1}{\lambda+2\mu}\log\left(\frac{V_0}{m_0^2\chi_0^2}\frac{\lambda}{\mu}\right)}_{\phi_0}-\log\left(\frac{\chi}{\chi_0}\right)^{\frac{2}{\lambda+2\mu}}\;.
\end{equation} 
We can now write
\begin{align}
    m_\chi&=m_0e^{\mu\phi_0}=m_0\left(\frac{V_0}{m_0^2\chi_0^2}\frac{\lambda}{\mu}\right)^{\frac{\mu}{\lambda+2\mu}},\\    \Lambda_0 &=V_0e^{-\lambda\phi_0}=V_0\left(\frac{V_0}{m_0^2\chi_0^2}\frac{\lambda}{\mu}\right)^{-\frac{\lambda}{\lambda+2\mu}}\label{eq: ccrunheavy},\\
    H_0&=\frac{\kappa_d}{\sqrt{(d-1)(d-2)}}\sqrt{2\Lambda_0+m_\chi^2\chi_0^2}\notag\\
    &=\sqrt{\frac{\lambda+2\mu}{\lambda(d-1)(d-2)}}\kappa_d m_\chi|\chi_0|=\sqrt{\frac{2\mu+\lambda}{2\mu(d-1)(d-2)}}\kappa_d\Lambda_0^{1/2},\\
    m_{\rm eff}^{(\phi)}& =\kappa_d^{-1}\sqrt{V_{\rm eff}''(\phi)|_{\phi_0}}=\sqrt{2\lambda\mu(d-1)(d-2)} H_0\;\label{eq.mPhi2}.
\end{align}
Again, the energy densities coming from the potential and the heavy scalar are comparable, as $\frac{\Lambda_0}{\rho_\chi}=\frac{\mu}{\lambda}=\mathcal{O}(1)$. We are left with the following equation of motion for the heavy scalar:
\begin{equation}
    \Ddot{\chi}+m_\chi^2\left(\frac{\chi}{\chi_0}\right)^{-\frac{4\mu }{\lambda+2\mu}}\chi=0\;,
\end{equation}
which we can solve in a $\mathcal{O}\left(m_\chi^2(t-t_0)^2\right)$ expansion as 
\begin{subequations}
\begin{align}
\kappa_d(\phi(t)-\phi_0)&=\frac{m_\chi^2}{\lambda+2\mu}(t-t_0)^2+\frac{m_\chi^4}{6}\frac{\lambda-4\mu}{\lambda^2-4\mu^2}(t-t_0)^4+\mathcal{O}(m_\chi(t-t_0))^6,\\
\chi(t)&=\chi_0\left\{1-\frac{m_\chi^2}{2}(t-t_0)^2+\frac{m_\chi^4}{24}\frac{\lambda+2\mu}{\lambda-2\mu}(t-t_0)^4+\mathcal{O}(m_\chi(t-t_0))^6\right\}\;.
\end{align}
\end{subequations}
    
\begin{figure}[H]
\begin{center}
\begin{subfigure}[b]{0.7\textwidth}
\center
\includegraphics[width=\textwidth]{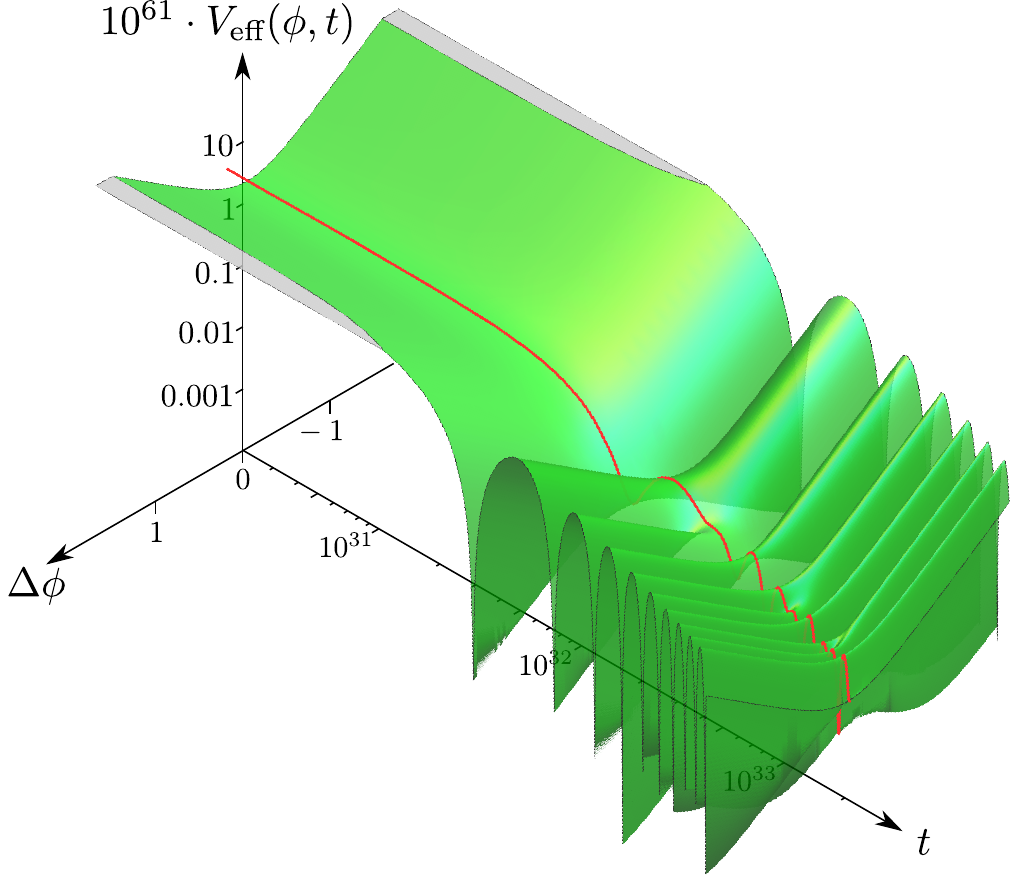}
\caption{Light and heavy states.} \label{f.LightHeavy3D}
\end{subfigure}
\begin{subfigure}[b]{0.7\textwidth}
\center
\includegraphics[width=\textwidth]{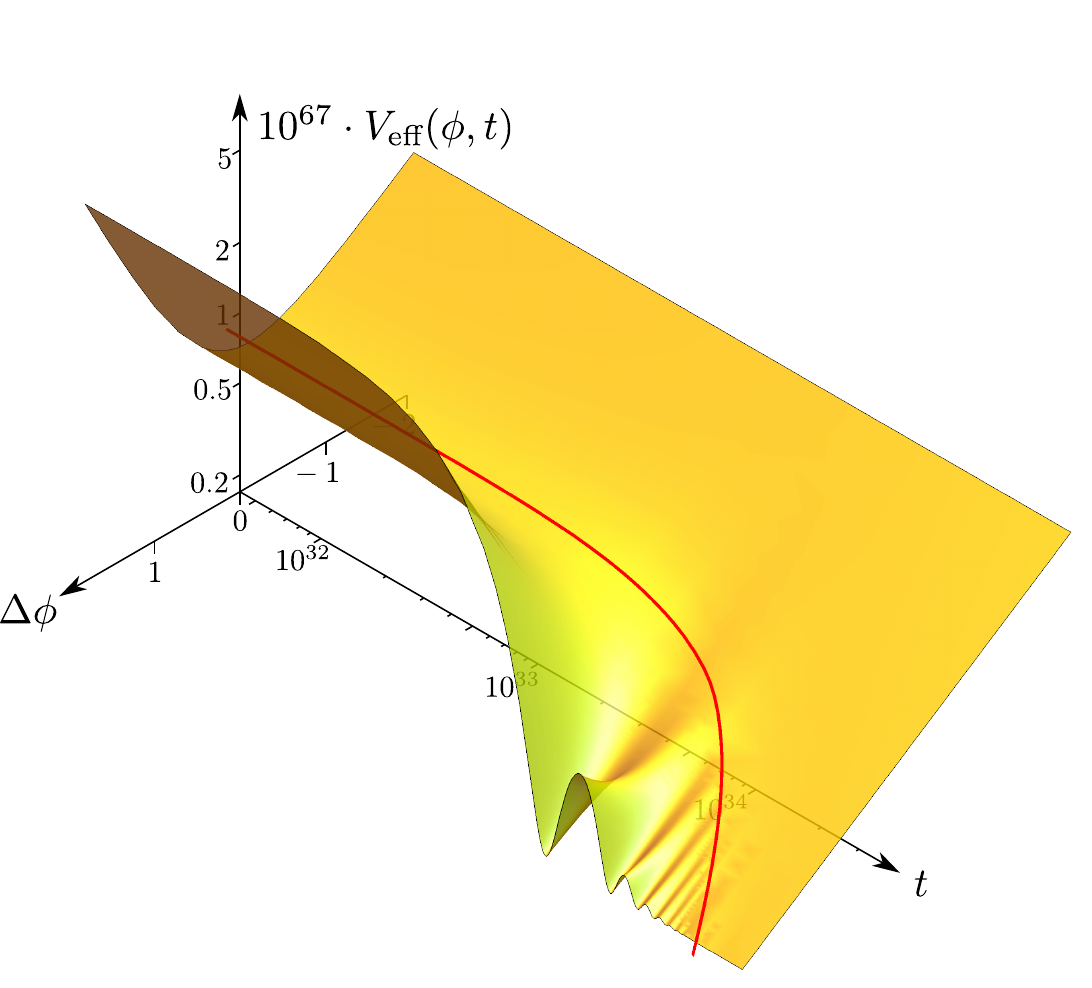}
\caption{Runaway potential and heavy states.} \label{f.PotHeavy3D}
\end{subfigure}
\caption{Time and moduli dependence of the effective potential $V_{ \rm eff}(\phi,t)$ for the two cases studied, with the trajectory carried out by the scalar $\phi$ in red. The parameters and initial conditions used are the same as in Figures \ref{f.PotLightPlots} and \ref{f.PotHeavy}, with all magnitudes given in Planck units.
\label{f.fullPot}}
\end{center}
\end{figure}

The quasi-dS phase ends after a period $t_{\rm dS}\approx \sqrt{2}m_\chi^{-1}$, after which the $\chi$ scalar starts oscillating and $\phi$ ceases to be fixed. The total number of $e$-folds during this transient era is given by
\begin{equation}
    N\approx t_{\rm dS}H_0=\sqrt{\frac{2(\lambda+2\mu)}{\lambda(d-1)(d-2)}}\kappa_d|\chi_0|\;.\label{eq: efoldsrunheavy}
\end{equation}
Same as before, we demand $\kappa_d|\chi_0|\lesssim 1$, which in turn results in
\begin{equation}
    N\leq \sqrt{\frac{2(\lambda+2\mu)}{\lambda(d-1)(d-2)}}\;.
\end{equation}
Assuming $\lambda>\frac{2}{\sqrt{d-2}}$ and $\mu\leq \sqrt{\frac{d-1}{d-2}}$ per the Swampland bounds, this becomes
\begin{equation}\label{e.NmaxHeavyPot}
    N\leq \sqrt{\frac{2(1+\sqrt{d-1})}{(d-1)(d-2)}}\;,
\end{equation}
which is saturated for $\lambda=\frac{2}{\sqrt{d-2}}$ and $\mu=\sqrt{\frac{d-1}{d-2}}$. Figures \ref{f.PotHeavy} and \ref{f.PotHeavy3D} show an example of numerical solutions to these equations.

\begin{figure}[h]
    \centering
    \includegraphics[width=0.75\linewidth]{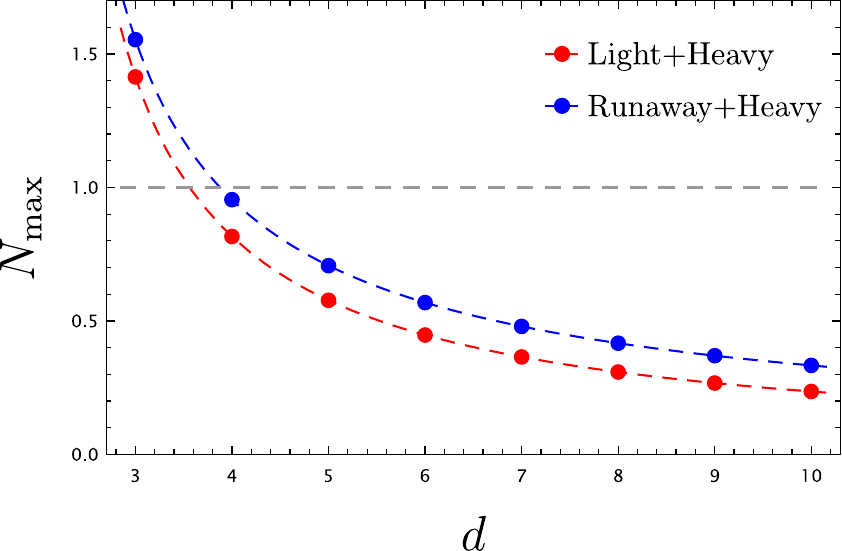}
    \caption{Comparison between the two upper bounds in the number $N$ of $e$-folds allowed by the two mechanisms, respectively a light and heavy tower, and a runaway potential and a heavy tower, in terms of the spacetime dimension $d$. Note that for $d\geq4$ it is not possible to obtain $N>1$ with subplanckian values for the scalar fields.}
    \label{f.comp}
\end{figure}

 Figure \ref{f.comp} shows the upper bounds for the number of $e$-folds in each dimension $d$ in the different constructions discussed. In both cases the obtained bounds are consistent with the TCC by checking that $t_{\rm dS}$ is short enough, as per \eqref{e.TCC lifetime}:
\begin{equation}\label{eq.TCC again}
    t_{\rm dS}\leq \frac{1}{H_0}\log\left(\frac{M_{\rm Pl,d}}{H_0}\right)\Longrightarrow N\approx t_{\rm dS} H_0\leq\log\left(\frac{M_{{\rm Pl},d}}{H_0}\right)\;.
\end{equation}
Using the expressions obtained along this section we can rewrite \eqref{eq.TCC again} as
\begin{equation}
    N_{\rm max}+\log N_{\rm max}\lesssim -\log\left(\frac{m_\chi}{M_{{\rm Pl}, d}}\right)-\log(\kappa_d|\chi_0|).
\end{equation}
This is satisfied with room to spare for subplanckian $\chi_0$, $\psi_0$, $m_\chi$ and $M_\psi$, since $N_{\rm max}\sim\mathcal{O}(1)$.

A difference with the two-tower scenario is that the value for  $w_\phi$ increases after the quasi-dS era, unlike what happened in Section \ref{s.light and heavy}. The reason is that the potential $V(\phi)$ does not dilute away, which makes $\phi$ pick up more velocity than in the previous scenario.
\vspace{0.8em}

A key point to consider for both scenarios is the scale at which the scalar field $\phi$ stabilizes. In both models the effective mass of this field, $m_{\rm eff}^{(\phi)}\sim H_0$, see \eqref{eq.mPhi1} and \eqref{eq.mPhi2}, could conflict with experimental constrains on fifth forces \cite{Adelberger:2003zx}, potentially invalidating the model. This issue is shared with quintessence scenarios (though in those cases the moduli are not stabilized), and can potentially be addressed by the \emph{traditional} chameleon mechanism \cite{Khoury:2003aq,Khoury:2003rn} that would allow the light scalar to evade Solar System tests.

\subsection{Generalization to several scalars\label{S.SeveralScalars}}
We will close this Section with a short discussion about how our setup generalizes to constructions with more than one modulus. We will collectively refer to these by $\vec\phi$ and also allow for a non-canonical moduli space metric $\mathsf{G}_{ab}$. In order to stabilize $\Vec{\phi}$ at some point $\vec\phi_0$, we need that the gradient of the effective potential vanishes at said position in $\mathcal{M}$, $\vec\nabla V_{\rm eff}(\vec\phi_0,t)=0$, where the derivatives are taken with respect to the moduli.

Working in an asymptotic region of the moduli space $\mathcal{M}$, we expect both the potential and the contributions of towers to depend exponentially on (locally) canonically normalized moduli:
\begin{equation}
    V_{\rm eff}(\vec{\phi},t)=\sum_{J}V_J(\vec\phi)+\frac{1}{2}\sum_{I}m_I(\vec\phi)\chi_I^2(t)\;,
\end{equation}
Defining now the  ``de Sitter ratios'' $\Vec{\mu}_J$ \cite{Calderon-Infante:2022nxb} of the potential and scalar charge-to-mass ratio vectors $\Vec{\zeta}_I$ \cite{Calderon-Infante:2020dhm,Etheredge:2022opl,Etheredge:2023odp} of the states as
\begin{equation}
    \mu^a_J(\Vec{\phi}_0)=-\delta^{ab}\mathsf{e}_b^i\partial_{\phi^i}\log \left(\frac{V_J(\vec\phi_0)}{M_{{\rm Pl},\, d}^d}\right),\qquad \zeta^a_I(\Vec{\phi}_0)=-\delta^{ab}\mathsf{e}_b^i\partial_{\phi^i}\log \left(\frac{m_I(\vec\phi_0)}{M_{{\rm Pl,}\, d}}\right)\;,
\end{equation}
where $\mathsf{e}_b^i$ are inverse vielbein of the moduli space metric, $\mathsf{G}^{ij}=\mathsf{e}_a^i\mathsf{e}_b^j\delta^{ab}$. If $\hat{\tau}$ is the normalized unit tangent vector to some asymptotic trajectory, the exponential rates of both the potential and the mass along said trajectory, $V_J(\Delta \phi)\sim e^{-\lambda_J\Delta\phi}$ and $m_I(\Delta \phi)\sim e^{-\alpha_I\Delta \phi}$, are locally given by
\begin{equation}
    \lambda_J=\hat{\tau}\cdot\Vec{\mu}_J,\quad \alpha_I=\hat{\tau}\cdot\Vec{\zeta}_I\;. 
\end{equation}

\begin{figure}[H]
\begin{center}
\begin{subfigure}[b]{0.45\textwidth}
\center
\includegraphics[width=\textwidth]{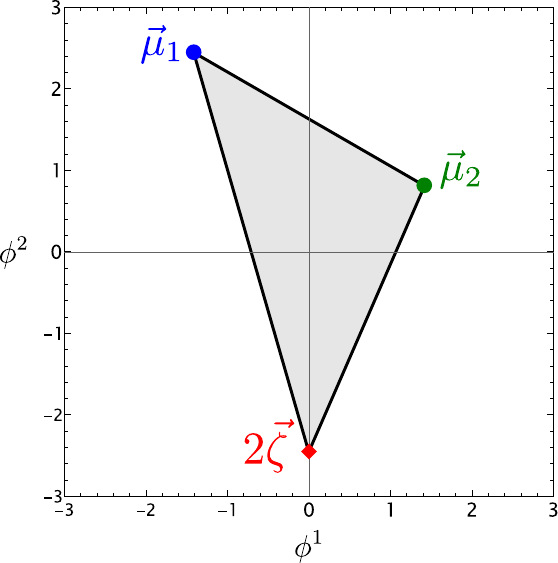}
\caption{Stabilizing effective potential} \label{f.CHa}
\end{subfigure}
\begin{subfigure}[b]{0.45\textwidth}
\center
\includegraphics[width=\textwidth]{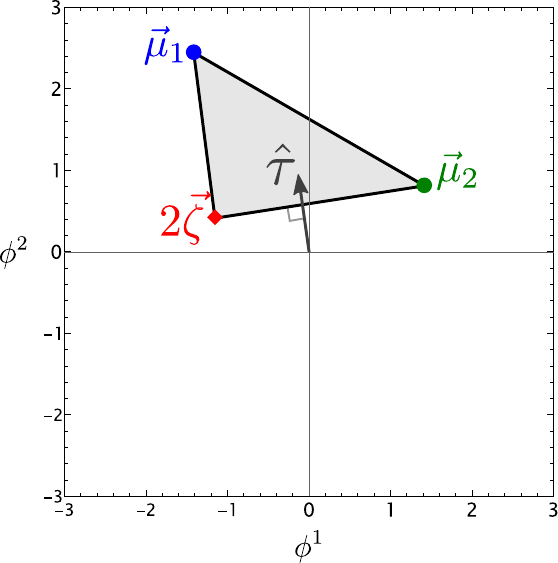}
\caption{Non-stabilizing effective potential.} \label{f.CHb}
\end{subfigure}

\caption{Convex hulls associated to the effective toy potentials given by a two-term potential and a (heavy) tower, respectively $V_{\rm eff}(\phi^1,\phi^2,t)={\color{blue}V_1e^{\sqrt{2}\phi^1-\sqrt{6}\phi^2}}+{\color{dark-green}V_2e^{-\sqrt{2}\phi^1-\sqrt{\frac23}\phi^2}}+{\color{red}\frac{m_0^2}{2} e^{\sqrt{6}\phi^2}\chi^2(t)}$ in \ref{f.CHa}  and  $V_{\rm eff}(\phi^1,\phi^2,t)={\color{blue}V_1e^{\sqrt{2}\phi^1-\sqrt{6}\phi^2}}+{\color{dark-green}V_2e^{-\sqrt{2}\phi^1-\sqrt{\frac23}\phi^2}}+{\color{red}\frac{m_0^2}{2} e^{\frac{2}{\sqrt{3}}\phi^1-\frac{1}{\sqrt{6}}\phi^2}\chi^2(t)}$ in \ref{f.CHb}. While \ref{f.CHa} features an effective minimum, \ref{f.CHb} does not, and thus the scalars move in the direction $\hat{\tau}$ determined by the leading terms, with $\phi^2\sim -2 \left(\sqrt{2}+\sqrt{3}\right) \phi^1>0$. \label{f.CH}
 }
\end{center}
\end{figure}

Following \cite{Calderon-Infante:2022nxb}, the condition $\vec\nabla V_{\rm eff}(\vec\phi_0,t)=0$ translates into the geometric requirement that the convex hull spanned jointly by the vectors of the potential and the towers contains the origin,
\begin{equation}\label{e.CH}
    \vec{0}\in{\rm Hull}_{\Vec{\phi}_0}\left(\{\Vec{\mu}_J\}_J\cup\{2\vec\zeta_I\}_I\right)\;.
\end{equation}
Here, the factor of two before $\vec\zeta_I$ takes into account that masses appear squared in the effective potential. An illustration appears in Figures \ref{f.CHa} and \ref{f.CHb}. This is equivalent to requiring that the scalar forces of the different terms on the different moduli balance each other, resulting in stabilization along any possible limit starting at $\vec\phi_0$. As each species dilutes, $\chi_I\to 0$, the convex hull \eqref{e.CH} evolves accordingly, with terms no longer contributing to the effective potential ceasing to be considered. Once it ceases to contain the origin, the scalars start moving with an initial velocity pointed towards the point in the convex hull closest to the origin \cite{Calderon-Infante:2022nxb, Shiu:2023fhb}, as illustrated in Figure \ref{f.CHb}. 

In general, for $K$ moduli one would need at least $K+1$ terms in the effective potential to achieve a minimum, with the quasi-dS epoch lasting $t_{\rm dS}=O(\min\{m_{I,0}^{-1}\}_I)$. As the stabilization mechanism results in all of the terms in the effective potential contributing in a similar way to the energy density, one would expect
\begin{equation}
    N\approx t_{\rm dS}H_0\approx \kappa_d\sqrt{\frac{2}{(d-1)(d-2)}}\frac{\sqrt{\sum_i\rho_{i,0}}}{\max\{m_{I}\}_I}\lesssim \mathcal{O}(\sqrt{K})\;,
\end{equation}
so that including more scalars (and thus more terms in the effective potential) in principle can allow for a higher number of $e$-folds, potentially larger than $\mathcal{O}(1)$. This is the same phenomenon that was observed in early Swampland discussions of axion WGC, where a setup with $K$ axions could enhance field range by a factor of $\sqrt{K}$ \cite{Bachlechner:2014gfa,Bachlechner:2015qja,Hebecker:2015rya,Montero:2015ofa,Baume:2016psm}, and we believe it should be explored further. In any case, as $K$ increases so does the difficulty in obtaining a controlled set of massive states and potential terms that behave in an appropriate way; see Section \ref{section 4: examplecitos} for our attempts.

\section{Range of validity of the EFT}\label{s: validity EFT}
In the previous Section we have illustrated how, by harnessing the power of the tower of states generically present in String Theory, it is possible to produce models sustaining accelerated expansion for $\mathcal{O}(1)$ $e$-folds. The precise number of $e$-folds that one can achieve is determined by the initial value $\vert \chi_0\vert$ that the scalar parameterizing the heavy field(s) can attain. $\mathcal{O}(1)$ number of $e$-folds corresponds to $\vert \chi_0\vert\sim \kappa_d^{-1}$. This poses a challenge, as one expects that neglected terms in the potential suppressed by powers of the Planck mass become important around these values. Furthermore, the scalar fields such as $\phi$ are the usual moduli of String Theory, and their effective action and couplings are well understood. By contrast, the scalar $\chi$ is a more exotic, emergent field, which accounts for the density of heavy particles present in our cosmology. The quadratic potential for $\chi$ used in the previous Section amounts to assuming that these particles are non-interacting, but of course, this is just an approximation. We need to know what is the range of validity of this approximation, in order to see if $|\chi_0|\sim \kappa_d^{-1}$ can be achieved to begin with. In this Section we describe the physical interpretation and interactions for $\chi$, and how these limit the allowed field range for this massive field.

First, consider the exactly free theory of a massive $\chi$ field coupled to gravity in some homogeneous state. In a classical picture,  a gas of massive particles of mass $m_\chi$ behaves as dust, so we can characterize this state by its number density $N$, which in four dimensions (the following discussion can be generalized for arbitrary $d$) has units of length$^{-3}$. The energy density of such a state is
\begin{equation}\rho_\chi= N m_\chi\;.\label{rho99}\end{equation}
On the other hand, from the canonical expression for the stress-energy tensor of a free scalar field, we know that 
\begin{equation}\rho_\chi\sim \frac{m_\chi^2}{2}\chi^2\;,\label{op1}\end{equation}
and so equating with \eq{rho99} gives
\begin{equation}\chi^2\sim \frac{2N}{m_\chi}.\label{op2}\end{equation}
One first limit comes from imposing that the density is subplanckian,
\begin{equation} N\lesssim M_{\rm Pl,4}^3\quad\Rightarrow\quad \chi \lesssim \left(\frac{M_{\rm Pl,4}}{m_\chi}\right)^{\frac12}\, M_{\rm Pl,4}.\end{equation}
This, however, still allows for very transplanckian field ranges,  as long as $m_\chi<M_{\rm Pl,4}$, which is always the case in our scenario.

However, the field range is significantly smaller if we allow interactions between $\chi$ particles. Consider a $\lambda \chi^4$ interaction. In this case, \eq{op1} is replaced by 
\begin{equation}\rho_\chi\sim \frac{m_\chi^2}{2}\chi^2+\frac{\lambda}{4!}\chi^4,\end{equation}
and for $\lambda\sim \mathcal{O}(1)$ the requirement that the energy density is subplanckian already forces us to have $|\chi|\lesssim M_{\rm Pl,4}$. In particular, we must have
\begin{equation} \vert \chi_0\vert^2\lesssim \frac{m_\chi}{\sqrt{\lambda}},\end{equation}
in order to suppress the effects of quartic and higher-order couplings in the potential. The value $\chi\sim m_\chi$ corresponds to a number density $N\sim m_\chi^{-3}$.  $m_\chi^{-1}$, precisely the Compton wavelength of the particle \cite{PhysRev.21.483}; picturing each individual $\chi$ particle as a fuzzy cloud of radius $\sim m_\chi^{-1}$, the above tells us that short-range interactions (such as $\chi^4$, which corresponds to a $\delta^{(3)}(\vec{r})$ potential between the $\chi$ particles \cite{TongQFT}) become important when the quantum clouds of each particle start touching one another.

For a 
\begin{equation}c_n\frac{\chi^{2n+4}}{\Lambda^{2n}}\end{equation}
term in the potential, suppressed by some power of a UV scale $\Lambda$, we must have
\begin{equation} \vert \chi_0\vert\lesssim  \frac{\Lambda}{c_n^{\frac{1}{2n+2}}}\left(\frac{m}{\Lambda}\right)^{\frac{1}{n+1}},\label{higherint}\end{equation}
which means that $\vert \chi_0\vert$ should be significantly smaller than the fundamental cutoff of the theory. This corresponds to the species scale which, in any setup with large extra dimensions, is strictly smaller than the Planck scale \cite{Dvali:2007hz,Dvali:2007wp,Dvali:2008ec}. 

As illustrated in the previous Section, we need $|\chi_0|\sim M_{\rm Pl,4}$ to have accelerated expansion for an $\mathcal{O}(1)$ number of $e$-folds.  Therefore, we need to suppress the interactions of the $\chi$ particle. In particular, we need to have 
\begin{equation}
\lambda\lesssim\frac{m_\chi^2}{M_{\rm Pl,4}^2},\qquad c_n\lesssim\frac{\Lambda^{2n} m_\chi^2}{M_{\rm Pl,4}^{2n+2}}\quad\text{for }n>0\;,
\label{suppr}
\end{equation}
if we are to have a Planckian field range for $\chi$. Realizing \eqref{suppr} is one of the most challenging requirements of the scenario, and it is similar to the challenges faced in constructing models of inflation where a scalar runs for a moderately large field range. We will discuss in Section \ref{section 4: examplecitos} how the suppression can be achieved; the idea is that we always live close to an asymptotic limit in moduli space which suppresses interactions (for instance, because there is a small overlap between wavefunctions of states in the large volume limit, etc). Possible caveats are also discussed.

Finally, we also need to be careful if the $\chi$ particle is subject to long-range forces other than those generated by the scalar $\phi$, as the energy density of the long-range force also contributes to the cosmology. In the examples below we will focus on massive states without long-range forces, so that this problem is averted. 

\section{Attempts at String theory embeddings \label{section 4: examplecitos}}
In the previous sections we have outlined how a scalar potential involving towers of fields of the kind typically obtained in String Theory can yield temporary phases of accelerated expansion, which may be compatible with our Universe. In this section we discuss stringy embeddings of these scenarios. The common ingredient is the presence of a tower of states (or at least, a single field) whose effective mass \emph{increases} as $\phi$ grows. Even though it is increasing, the mass of this object must be extremely light, since it contributes an $\mathcal{O}(1)$ amount to the vacuum energy density during the phase of accelerated expansion. For instance, from expression \eq{eq: runheavy}, we get that 
\begin{equation}V_0\sim   \kappa_d^2 m_0^2 \vert\chi_0\vert^2.\end{equation}
Setting $V_0$ to be the vacuum energy density, as well as assuming $\vert\chi_0\vert^2$ not too small, $m_0$ comes about to Hubble scale mass. In the case of the light and heavy towers of states, the masses of both sets of fields should be roughly the same when $\phi=\phi_0$.

This means that, for our mechanism to work, we need to locate points in string moduli spaces where heavy and light towers are roughly the same mass. For supersymmetric moduli spaces, this  happens typically close to \emph{desert points} \cite{Long:2021jlv,vandeHeisteeg:2022btw,vandeHeisteeg:2023ubh}, where e.g. a $U(1)$ coupling is of order 1, so electric and magnetic states have similar masses. On the other hand, at desert points it is precisely the case that the towers of states are already very heavy \cite{Long:2021jlv}, with masses close to the species scale. Similarly, if a potential is generated, one expects it to be of order the species scale as well.

Another consideration involves the stability of the heavy tower. A tower of states which is becoming heavier is also becoming more and more strongly coupled, and will decay into lighter modes unless protected by some exact symmetry. Fortunately, as explained in Section \ref{S.CosmoCh}, the lightest state of the tower is enough to produce the desired effect, as long as it gets heavy sufficiently quickly and is stable. We will therefore look for towers protected by conserved charges (continuous or discrete).

The challenge is therefore to find stringy setups where there are states both becoming light and heavy, of comparable mass, and such that this mass is small compared to the Planck and species scales. We also need to ensure that the interactions of the heavy tower are suppressed as explained in Section \ref{s: validity EFT}. We now discuss some examples schematically. Unfortunately, issues remain in all of them, precluding a full stringy embedding of the Cosmological Chameleon scenario; we nevertheless hope that these issues can be overcome in the future, and that our schematic analysis here can serve as a first step.

\subsection{Calabi-Yau close to an SCFT point}
For our first scenario, consider M-theory on a background of the form $\mathbb{M}^4\times X_3\times S^1$, where $X_3$ is a compact Calabi-Yau threefold. As a toy model, we will assume that all moduli of $X_3$ have been stabilized, possibly in a non-supersymmetric way, or via the use of dualities like in \cite{Gkountoumis:2023fym}. Having done this, we will assume the single modulus present in the system to be $R$, the size of the $S^1$.  Different states can emerge in this configuration, ranging from Kaluza-Klein bulk modes to extended objects (membranes or their magnetic duals) wrapping cycles in the inner dimensions.  We will focus on a heavy state arising from an M5 membrane wrapping a four-cycle $\Sigma_4 \in H_{2,2}(X_3,\mathcal{R})$ and $S^1$. This state is seen as a low energy particle in 4d. Alternatively, it can be thought of as a winding mode of a string in 5d.  Furthermore, we assume that the cycle $\Sigma_4$ is stabilized close to, but not exactly at a SCFT point of zero size Vol$(\Sigma_4)\xrightarrow{}0$, in $M_{\rm Pl,11}$ units.
Up to now, three different moduli fields arise, the total volume $\mathcal{V}_X$ of $X_3$ (which we also assume to be stabilized), that of the four-cycle $Z\equiv \text{Vol}(\Sigma_4)$ and the radius $R$ of the $S^1$ (in $M_{\rm Pl,5}$ units). The mass of the particle takes the form

\begin{equation}
    M_R = Z\,R \,\,M_{\rm Pl,11} = \hat{Z}\sqrt{R} \,\,M_{\rm Pl,4}\;,
\end{equation}
where we redefine $\Hat{Z}=Z\mathcal{V}_X^{-1/2}$.
The $S^1$ radius can be canonically normalized as
\begin{equation}
    \rho =\sqrt{\frac{d-1}{d-2}} \log\left(\frac{R}{R_0}\right)= \sqrt{\frac{3}{2}} \log\left(\frac{R}{R_0}\right)\;,
\end{equation}
thus arriving at a term $\frac{1}{2}\hat{Z}^2R_0e^{\sqrt{\frac{2}{3}}\rho}\chi^2M_{\rm Pl,4}^2$ in the effective potential, with $\chi$ the vev of such winding mode.

For the part of the effective potential becoming light with $\rho\to\infty$, we may consider simply the KK tower. However, another option is the Casimir energy that may emerge from loop corrections after circle compactifications. As it is well known from the literature \cite{Arkani-Hamed:2007ryu}, Casimir energies in 4d have a radius dependence of the form $V_{\rm Casimir}\sim R^{-4}$. Putting it all together, the effective potential under consideration has the form
\begin{equation}
    V_{\rm eff}(r,\chi) = V_0\, e^{-4\sqrt{\frac{2}{3}}\rho}  + \frac{M_{\rm Pl,4}^2}{2}\hat{Z}^2R_0 e^{\sqrt{\frac{2}{3}}\rho}\chi^2\;.
\end{equation}

We arrive at an expression of the form \eqref{eq: runheavy} up to stabilizing the modulus $\hat Z$. As shown in \eqref{eq: ccrunheavy}, $\hat Z$ must be very small to obtain a suppressed cosmological constant.  This is why it was assumed to be close to the SCFT point. As the string is strongly coupled, we expect it will decay, and only the lowest-lying winding mode will survive as a suitable candidate out of all the states in the tower.

We are finally in conditions to find the maximum number of $e$-folds allowed, using \eqref{eq: efoldsrunheavy}, as for this model we have $\mu=\frac{1}{\sqrt{6}}$ and $\lambda=4\sqrt{\frac{2}{3}}$. So,

\begin{equation}
    N \approx\frac{1}{2}\sqrt{\frac{5}{3}}\, \kappa_5 |\chi_0|\approx 0.6455\, \kappa_5 |\chi_0|
\end{equation}
This result is borderline compatible with our universe, as we measure roughly $0.5$ $e$-folds since the beginning of the current accelerated era of expansion. However, these numbers are obtained only when $\chi_0\sim M_{\rm Pl,4}$, where the corrections discussed in Section \ref{s: validity EFT} to the simple action we considered become important. In short, the above shows that engineering a Cosmological Chameleon near an SCFT point is not obviously impossible, but there are many issues to overcome.

\subsection{Multifield set-ups and difficulties encountered\label{s.examples multifields}}
We now consider examples with more than one moduli, and comment on some problems found when trying to obtain stringy embeddings for these. As an illustrative example, we will consider a 4d $\mathcal{N}=1$ EFT obtained from type IIA string theory compactified on a Calabi-Yau orientifold $X_3$ with a single K\"ahler modulus, $h^{1,1}(X_3)=1$. We will assume that all possible complex structure moduli are stabilized via a combination of fluxes and orientifolds, similar to e.g. \cite{Grimm:2004ua,DeWolfe:2005uu,Camara:2005dc}. As such the two non stabilized moduli will consist on the Type IIA 10d dilaton $\phi$ and the manifold volume $V_X$, measured in 10d Planck units. The associated  moduli space metric in the 4d Einstein frame is diagonal and given by $
    \mathsf{G}_{\phi\phi}=\frac{1}{2},$ and $\mathsf{G}_{V_XV_X}=\frac{2}{3V_X^2}$, through which we define the following canonically normalized moduli:
\begin{equation}
    \hat{\phi}=\frac{\phi}{\sqrt{2}},\qquad\rho=\sqrt{\frac{2}{3}}\log V_X\;.
\end{equation}
We further assume the existence of 2- and 4-cycles in $X_3$. As the complex structure moduli are stabilized, their volume will scale parametrically with $V_X$, $V_2=v_2 V_X^{1/3}$ and $V_4=v_4 V_X^{2/3}$, with $v_2,\,v_4>0$ some constants. As candidates to heavy states we will consider wrapped D2 and D4 branes, with tension \cite{Polchinski:1998rr} $\tau_p\sim e^{-\phi}(\alpha')^{\frac{p+1}{2}}\sim e^{\frac{p-3}{4}\phi}M_{\rm Pl\,10}^{p+1}$, so that
\begin{subequations}
    \begin{align}
    \frac{m_{\rm D2}}{M_{{\rm Pl,}4}}&\sim \frac{M_{\rm Pl,10}}{M_{{\rm Pl,}4}}\tau_2V_2\sim \exp\left\{-\frac{\hat{\phi}}{2\sqrt{2}}-\frac{{\rho}}{2\sqrt{6}}\right\}\\
    \frac{m_{\rm D4}}{M_{{\rm Pl,}4}}&\sim \frac{M_{\rm Pl,10}}{M_{\rm Pl,4}}\tau_4V_4\sim \exp\left\{\frac{\hat{\phi}}{2\sqrt{2}}+\frac{{\rho}}{2\sqrt{6}}\right\}\;,
\end{align}
\end{subequations}
being electric-magnetic dual states. From the analysis of Section \ref{S.SeveralScalars} we have that 
\begin{equation}
    2\Vec{\zeta}_{\rm D2}=\left(\frac{1}{\sqrt{2}},\frac{1}{\sqrt{6}}\right),\qquad 2\Vec{\zeta}_{\rm D4}=\left(-\frac{1}{\sqrt{2}},-\frac{1}{\sqrt{6}}\right)\,,
\end{equation}
with the segment joining them crossing the origin. One will need then the $\vec\mu$ vector associated with the terms in the 
(flux) potential $V(\hat{\phi},\rho)$ to be such that the resulting convex hull contains the origin. As we will show now, this is not as easy as it seems.

\begin{figure}[h]
\begin{center}
\begin{subfigure}[b]{0.49\textwidth}
\center
\includegraphics[width=\textwidth]{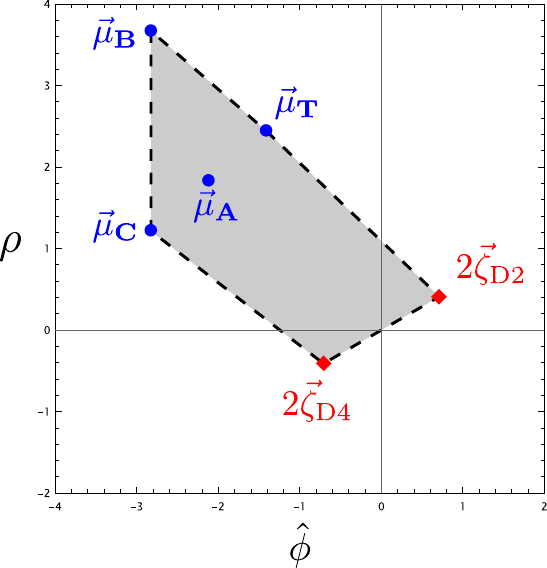}
\caption{Effective potential \eqref{e.DGKTpot}} \label{f.DGKTpot}
\end{subfigure}
\begin{subfigure}[b]{0.49\textwidth}
\center
\includegraphics[width=\textwidth]{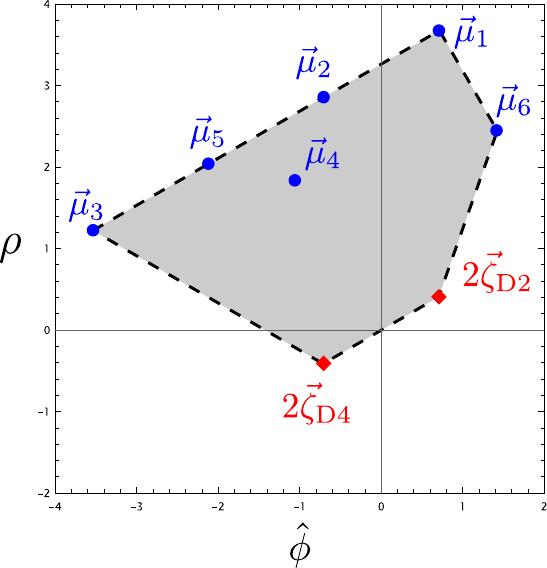}
\caption{\cite{Calderon-Infante:2022nxb}+D2+D4 effective potential.} \label{f.OTHERpot}
\end{subfigure}

\caption{Representation of the convex hulls associated with the effective potentials resulting from \eqref{e.DGKTpot} and \cite{Calderon-Infante:2022nxb}.  In both cases wrapped D2 and D4 branes are used as heavy states. Irrespective of the terms that are turned off or on, none of the convex hulls contain the origin \emph{inside} them. At most $\vec 0$ can be contained in the edge spanned by $2\vec\zeta_{\rm D2}$ and $2\vec\zeta_{\rm D4}$, which would stabilize said direction with the scalars moving in the perpendicular one. \label{f.CHpots}
 }
\end{center}
\end{figure}

We first consider the massive IIA potential with Romans mass and O6/D6 sources \cite{Grimm:2004ua,DeWolfe:2005uu,Camara:2005dc}, which in terms of canonically normalized scalars is given by
\begin{equation}\label{e.DGKTpot}
    V_{\rm IIA}(\hat{\phi},\rho)= \mathbf{A} e^{\sqrt{2}\hat\phi-\sqrt{6}\rho}+\mathbf{B}e^{2\sqrt{2}\hat\phi-3\sqrt{\frac{3}{2}}\rho}+\mathbf{C}e^{2\sqrt{2}\hat{\phi}-\sqrt{\frac32}\rho}+\mathbf{T}e^{\frac{3}{\sqrt{2}}\hat{\phi}+\frac{3}{2}\sqrt{\frac32}\rho}\;,
\end{equation}
where $\mathbf{A}$, $\mathbf{B}$, $\mathbf{C}$ and $\mathbf{T}$ are functions of the RR and NSNS flux quanta $F_0$, $H_3$ and $F_4^i$. $\mathbf{T}$ is the contribution coming from O6 planes and D6 branes. If one is looking for supersymmetric solutions, $\mathbf{T}$ has a negative value from the tadpole cancellation condition. Since we are interested in a running solution rather than a vacuum, we can take the four prefactors positive and independent from each other; this amounts to adding a number of D6 branes to cancel the tadpole, whose moduli stabilization we do not discuss (in any case, we expect them to just find a minimum at a suitable location in the Calabi-Yau, not affecting the discussion here). The resulting convex hull for $V_{\rm eff}(\hat{\phi},\rho, t)$ is depicted in Figure \ref{f.DGKTpot}. As all the $\mu$-vectors are located on the same side of the $2\vec\zeta_{\rm D2}$-$2\vec\zeta_{\rm D4}$ edge, the origin is never contained inside the convex hull and thus there is no effective minimum.

Other possibilities for $V(\hat\phi,\rho)$ can be found in \cite{Calderon-Infante:2022nxb}. For example, consider the following two combinations of the Calabi-Yau volume in string units and the type IIA 10d dilaton:
\begin{equation}
    s=e^{-\phi}\sqrt{V_s}=e^{-\phi/4}\sqrt{V_X}=s_0 e^{\frac{\hat{\phi}}{2\sqrt{2}}+\frac{1}{2}\sqrt{\frac{3}{2}}\rho} \qquad u=V_s^{1/3}=e^{\phi/2}V_X^{1/3}=u_0 e^{\frac{\hat{\phi}}{\sqrt{2}}+\frac{\rho}{\sqrt{6}}}\;.
\end{equation}
These definitions are motivated by the IIB mirror dual  \cite{Calderon-Infante:2022nxb}. Ignoring stabilization of the IIA complex structure moduli, mirror symmetry results in the following potential for $s$ and $u$:
\begin{align}\label{e.IIA pot}
    V&\propto\frac{A_{-4,-3}}{s^4u^3}+\frac{A_{-4,-1}}{s^4u}+A_{-4,1}\frac{u}{s^4}+A_{-4,3}\frac{u^3}{s^4}+\frac{A_{-2,-3}}{s^2u^3}+\frac{A_{-3,0}}{s^3}\notag\\ 
    &\equiv\mathbf{A_1}e^{-\frac{1}{\sqrt{2}}\hat{\phi}-3\sqrt{\frac{3}{2}}\rho}+
    \mathbf{A_2}e^{\frac{1}{\sqrt{2}}\hat{\phi}-\frac{7}{\sqrt{6}}\rho}+
    \mathbf{A_3}e^{\frac{3}{\sqrt{2}}\hat{\phi}-\frac{5}{\sqrt{6}}\rho}+
    \mathbf{A_4}e^{\frac{5}{\sqrt{2}}\hat{\phi}-\sqrt{\frac{3}{2}}\rho}\notag\\
    &\qquad+
    \mathbf{A_5}e^{-\sqrt{2}\hat{\phi}-\sqrt{6}\rho}+
    \mathbf{A_6}e^{-\frac{3}{2\sqrt{2}}\hat{\phi}-\frac{3}{2}\sqrt{\frac{3}{2}}\rho}\;,
\end{align}
where the prefactors depend on the flux quanta and can be turned off for particular choices (again see \cite{Calderon-Infante:2022nxb} for more details). The convex hull for $V_{\rm eff}(\hat{\phi},\rho, t)$ is depicted in Figure \ref{f.OTHERpot}. The same way as in the previous case, all the $\mu$-vectors, irrespective of the possibility that some of the might be turned off, lay at the same side of the $2\vec\zeta_{\rm D2}$-$2\vec\zeta_{\rm D4}$ segment, preventing an effective minimum from materializing.

The two above examples considered serve as an illustration on how complicated it is for the potential and heavy states to conspire in such a way that their convex hull contains the origin. Of course this is not definite evidence that Chameleon constructions are not possible when more than one modulus are considered--after all only a particular subset of type IIA constructions has been briefly studied--, but again points in the direction that finding such set-ups is not an easy feat.

\subsection{The conifold}

As our final example using Calabi-Yau compactifications, we consider the conifold \cite{Strominger:1995cz}. The difference with the previous constructions lies in the fact that states engineering the Cosmololgical Chameleon become light at a point in the \emph{bulk} of the moduli space, namely a \emph{conifold point}. As we shall see, the mass dependence of these states is not exponential with the canonically normalized modulus, which is an important difference from the rest of the constructions of this Section, which are asymptotic (they involve fields rolling to infinite distance).

Consider for this type IIB string theory compactified to $d=4$ on a CY threefold $X_3$. The resulting $\mathcal{N}=2$ theory contains $\{Z^I\}_{I=1}^{h_{1,2}(X_3)}$ complex structure moduli, consisting in the periods of the holomorphic 3-form $\Omega$. For simplicity, we will consider that all but one of the scalars are stabilized ($X_3$ volume $\mathcal{V}_X$ included), and take $Z$ as our complex modulus, with metric $\mathsf{G}_{Z\Bar{Z}}\sim \log|Z|^2$. Around the conifold point, $Z=0$, the corresponding 3-surface $\mathcal{B}\subset X_3$ degenerates to a point. D3-branes wrapping $\mathcal{B}$ have mass proportional to \cite{Strominger:1995cz}
\begin{equation}\label{e.coni D3}
    \frac{m_{\rm D3}}{M_{\rm Pl,4}}\sim |Z|,
\end{equation}
vanishing at the conifold point. According to section \ref{s.light and heavy}, this state corresponds to the light one. The heavy state in this configuration would correspond to the dual magnetic object, given by another D3-brane wrapping the dual three-cycle $\mathcal{A}$, where $[\mathcal{A}]\cdot[\mathcal{B}]=1$. This three-cycle $\mathcal{A}$ does not lie in the conifold geometry and carries global information of the Calabi-Yau manifold. Therefore the mass of the heavy state would be proportional to the volume of $\mathcal{A}$ and hence constant (having assumed $\mathcal{V}_X$ stabilised). This however results in light states not being needed, as heavy states already produce a term  $ \frac{1}{2}m_\chi^2\, \chi^2>0$ in the effective  potential, resulting in a de-Sitter vacuum. One concludes then that a 1-modulus cosmological Chameleon construction cannot be easily built around the conifold point.

\vspace{0.5cm}

This is not surprising and teaches us that for our model to be applicable, as we have seen in the previous examples and in this one, it is necessary to keep the global aspects of compactification under tight control. Finally, for our last example, we will try to put all these lessons together and use a construction with a single modulus.

\subsection{Hyperbolic compactifications of M-theory}\label{s: veryverytrusty}
The examples discussed so far have the basic problem that some moduli are left unstabilized, providing potentially very steep directions for the potential. To avoid this, it would be desirable to have a scenario where there is a single modulus -- the volume -- that is stabilized by the Cosmological Chameleon. To do this, we will consider M-theory compactified on a hyperbolic 7-manifold $\mathcal{H}_7$ with volume $\mathcal{V}_{\mathcal{H}}$ in 11d Planck units. The choice of a hyperbolic manifold is motivated by the fact that all moduli save $\mathcal{V}_{\mathcal{H}}$ are automatically stabilized \cite{Townsend:2003fx,Ohta:2004wk} (even after cusps, if they exist, are removed via Anderson's theory; the resulting manifold has an Einstein metric with suitable rigidity properties); we could do a similar construction with any other family of spaces where the only unstabilized modulus is the volume. Dimensional reduction by compactifying an 11d theory on $\mathcal{H}_7$ results in the following kinetic term in the 4d Einstein frame metric \cite{Castellano:2023jjt},
\begin{equation}
    S_{\rm 4d}\supseteq-\frac{1}{2\kappa_4^2}\int\diff^4x\sqrt{-g}\frac{9}{14}\left(\partial \log \mathcal{V}_\mathcal{H}\right)\;,
\end{equation}
which allows us to define the canonically normalized modulus
\begin{equation}
    \rho=\frac{3}{\sqrt{14}}\log\frac{\mathcal{V}_\mathcal{H}}{\mathcal{V}_{\mathcal{H},0}}\Longleftrightarrow \mathcal{V_\mathcal{H}}=\mathcal{V}_{\mathcal{H},0}e^{\frac{\sqrt{14}}{3}\rho}\;.
\end{equation}
Under the assumption of homogeneity, one can define $R_\mathcal{H}=\mathcal{V}_\mathcal{H}^{1/7}=R_{\mathcal{H},0}\exp\left\{\frac{1}{3}\sqrt{\frac{2}{7}}\rho\right\}$ as the characteristic scale of our internal space.

After identifying the relevant modulus, we turn our attention to the runaway potential. The (constant) negative curvature of $\mathcal{H}_7$ results in a positive term
\begin{equation}
    \frac{V_R(\rho)}{M_{\rm Pl,\,4}^4}=-\frac{\mathcal{V}_{\mathcal{H},0}}{2\mathcal{V}_{\mathcal{H}}^2}\int\diff^7 x\sqrt{-g_7}R_{g_7}=-\frac{1}{2}\mathcal{V}_{\mathcal{H},0}R_{g_7,0}\mathcal{V}_{\mathcal{H}}^{-9/7}=\frac{V_R(\rho_0)}{M_{\rm Pl,\,4}^4}\exp\left\{-3\sqrt{\frac{2}{7}}\rho\right\}\;.
\end{equation}

Then we consider mass terms arising from  $M2$ and $M5$ branes respectively wrapping 2- and 5- cycles  $\Pi_2,\,\Pi_5\subset\mathcal{H}_7$. An additional assumption is that $\mathcal{H}_7$ has torsion in its second and fifth homology group.\footnote{Note that for an oriented $n$-manifold without boundary $M$ we have that the free part of the homology is identified as $fH_k(M;\mathbb{Z})\simeq fH_{n-k}(M;\mathbb{Z})\simeq \mathbb{Z}^{b_k(M)}$, and the torsion part is $\tau H_k(M;\mathbb{Z})\simeq \tau H_{n-k-1}(M;\mathbb{Z})$. In the case $n=7$, 2- and 5-cycles are dual in their free parts (so we will have the same number of non-trivial 2- and 5-cycles) but not in their torsion. Having torsion in them requires also having torsion in 4- and 1-cycles, respectively.} We have not explicitly checked that hyperbolic 7-manifolds with such torsion exist; however, hyperbolic manifolds with torsion do exist in three dimensions \cite{porti2016reidemeister}, and so one could hope that torsional examples also exist in higher dimensions. We leave a more thorough study of hyperbolic 7-manifolds to future work. Their masses in 4d Planck units are
\begin{subequations}\label{e.M2M5mass}
    \begin{align}
    \frac{m_{\rm M2}(\rho)}{M_{\rm Pl,\,4}}&=\frac{M_{\rm Pl,\,11}}{M_{\rm Pl,\,4}}M_{\rm Pl,\,11}^2v_2R_\mathcal{H}^2=M_{\rm Pl,\,11}^2v_2\mathcal{V}_\mathcal{H}^{-3/14}=\frac{m_{\rm M2}(\rho_0)}{M_{\rm Pl,\,4}}e^{-\frac{\rho}{\sqrt{14}}}\\
    \frac{m_{\rm M5}(\rho)}{M_{\rm Pl,\,4}}&=\frac{M_{\rm Pl,\,11}}{M_{\rm Pl,\,4}}M_{\rm Pl,\,11}^5v_5R_\mathcal{H}^5=M_{\rm Pl,\,11}^5v_5\mathcal{V}_\mathcal{H}^{3/14}=\frac{m_{\rm M5}(\rho_0)}{M_{\rm Pl,\,4}}e^{\frac{\rho}{\sqrt{14}}}\;,
\end{align}
\end{subequations}
where we have defined ${\rm Vol}(\Pi_2)=v_2 R_{\mathcal{H}}^2$  and ${\rm Vol}(\Pi_5)=v_5 R_{\mathcal{H}}^5$ as the volume of the internal cycles. From the discussion around \eqref{e.initialEQ}, for similar initial vevs in the resulting fields, the initial masses are approximately equal, $m_{\rm M2}(\rho_0)\sim m_{\rm M5}(\rho_0)$, which translates to a relation between the cycle volumes $\frac{v_2}{v_5}=(R_{\mathcal{H},0}M_{{\rm Pl,\,11}})^3$. If we want to work in the large volume regime, then this translates to $v_2\gg v_5$. Again we leave for future work the question of whether one can find hyperbolic 7-manifolds with torsion 2- and 5-cycles with the required hierarchy in their cycle volumes.

Finally, there is another light tower, that of KK modes on the whole of $\mathcal{H}_7$. Their masses scale as
\begin{equation}
    \frac{m_{\rm KK}}{M_{\rm Pl,\,4}}=\frac{M_{\rm Pl,\,11}}{M_{\rm Pl,\,4}} V_{\mathcal{H}}^{-1/7}\sim e^{-\frac{3}{\sqrt{14}}\rho}\;.
\end{equation}
The fact that the KK tower is lighter than our wrapped membranes means that the setup is outside of the regime of validity of a traditional EFT framework, since all the states of the KK tower should be included in the EFT. More concretely, we would expect the existence of couplings between the KK modes and the $\psi$ and $\chi$ fields counting M2 and M5 brane states. These couplings could, in principle, significantly alter the conclusions of our analysis by making these scalars decay quicker than expected and shortening the accelerating phase.

In order to roughly estimate the value of these couplings, we will assume that for practical efforts, our manifold can be seen as a factorization $\mathcal{H}_7\simeq\mathcal{F}_p\times \mathcal{B}_{7-p}$, with $p=2,\,5$. We first wrap the M$p$ brane on $\mathcal{F}_p$, which in $11-p$ dimensions results in a particle with mass
\begin{equation}
    \frac{\mathsf{m}_{\text{M}p}}{M_{\text{Pl},11-p}}=\frac{M_{\rm Pl,11}}{M_{\text{Pl},11-p}} M_{\rm Pl,11}^pV_{\mathcal{F}}=M_{\rm Pl,11}^p V_{\mathcal{F}}^{\frac{p-8}{p-9}}\,,
\end{equation}
where $V_{\mathcal{F}}$ is the volume of $\mathcal{F}_p$ in $M_{\rm Pl,11}$ units. This results in the following terms for the $(11-p)$-dimensional Einstein action:
\begin{equation}
    S_{11-p}\supset\frac{1}{2}\int\diff^{11-p}x\sqrt{-g_{11-p}}\left\{\kappa_{11-p}^{-2}\mathcal{R}_{g_{11-p}}-\mathsf{m}_{{\rm M}p}^2\chi_p^2\right\}\;,
\end{equation}
where $\chi_p$ are the particles resulting from wrapping the branes. In order to compactify down to $d=4$, we split our metric as
\begin{equation}
    g_{11-p}=\left({g}_4\frac{\mathsf{V}_{\mathcal{B},0}}{\mathsf{V}_{\mathcal{B}}}\right)\otimes\left(h_{\mathcal{B}}\sum_{n=0}^{\infty}f_n^2\kappa_{11-p}^2\right)\,,
\end{equation}
where $\mathsf{V}_{\mathcal{B}}$ is the volume of $\mathcal{B}_{7-p}$ in $M_{\text{Pl},11-p}$ units and $\{f_n\}_{n=0}^\infty$ are the KK expansion of the internal metric over a constant volume-1 metric $h_\mathcal{B}$. We take $f_0=\mathsf{V}_{\mathcal{B}}^{\frac{1}{7-p}}\kappa_{11-p}^{-1}$, constant over $\mathcal{B}_{7-p}$. This allows us to obtain the $d=4$ Einstein action as
\begin{align}
   S_4&\supset\frac{1}{2}\int\diff^4x\sqrt{-g_4}\left\{\kappa_4^{-2}\mathcal{R}_{g_4}-\left(1+\frac{7-p}{2f_0^2\kappa_{11-p}^2}\sum_{n=1}^\infty\frac{\hat{f}_n^2}{M_{\rm Pl,4}^2}\right)m_{{\rm M}p}^2\hat\chi_p^2\right\} \notag\\
   &=\frac{1}{2}\int\diff^4x\sqrt{-g_4}\left\{\kappa_4^{-2}\mathcal{R}_{g_4}-m_p^2\hat{\chi}_p^2-\frac{7-p}{2\mathsf{V}_{\mathcal{B}}^{\frac{2}{7-p}}}\frac{m_{{\rm M}p}^2}{M_{\rm Pl,4}^2}\sum_{n=1}^\infty\hat\chi_p^2\hat{f}_n^2\right\}\;,
\end{align}
where now\footnote{Using that in $M_{\rm Pl,11}$ units $V_{\mathcal{B}}=\mathsf{V}_\mathcal{B}V_\mathcal{F}^{-\frac{p-7}{p-9}}$ one recovers expressions \eqref{e.M2M5mass} for $m_{{\rm M}p}$.} $m_{{\rm M}p}=\left(\frac{\mathsf{V}_{\mathcal{B},0}}{\mathsf{V}_{\mathcal{B}}}\right)^{1/2}\mathsf{m}_{{\rm M}p}$ and $\hat\chi_p=\mathsf{V}_{\mathcal{B},0}^{1/2}\chi_p$ are in the appropriate $M_{\rm Pl,4}$ units and we have considered the KK modes orthogonal and normalized in $M_{\rm Pl,4}$ units as
\begin{equation}
    \int_{\mathcal{B}}\diff^6x\sqrt{h_\mathcal{B}}f_n f_m\kappa_{11-p}^2=\delta_{nm}\hat{f}_n^2\kappa_{4}^2\;.
\end{equation}
This way we identify the (adimensional) quartic coupling $\lambda_p\hat{f}^2_n\hat\chi_p^2$ as
\begin{equation}
    \lambda_p=\frac{7-p}{2}\frac{m_{{\rm M}p}^2}{M_{\rm Pl,4}^2} \mathsf{V}_{\mathcal{B}}^{\frac{2}{p-7}}= \frac{7-p}{2}\frac{m_{{\rm M}p}^2}{M_{\rm Pl,4}^2}v_p^{\frac{4}{(p-7)(p-9)}}V_\mathcal{H}^{\frac{2}{p-9}}\;.
\end{equation}
Initially having $m_{M2}\sim m_{M5}$ translates in
\begin{equation}
    \frac{\lambda_2}{\lambda_5}=\frac{5}{2}v_2^{-\frac{27}{70}}V_{\mathcal{H},0}^{-1/2}\ll 1\;,
\end{equation}
so the M5 interacts more strongly with the geometry. Being $\lambda_p$ independent of the level $n$ of the KK modes, the $\chi_p-f_n$ couplings are important if
\begin{equation}
    \lambda_p\left(\sum_{n=1}^Nf_n^2\right)\chi_p^2\sim m_{{\rm M}p}^2\chi_p^2\Longleftrightarrow v_p^{\frac{4}{(p-7)(p-9)}}V_{\mathcal{H}}^{\frac{2}{p-9}}\sum_{n=1}^N\left(f_n\kappa_4\right)^2\sim 1\,,
\end{equation}
$N$ is the number of KK modes excited by the interaction with $\chi_p$, and for clarity reasons we do not write hats on top of $\chi_p$ and $f_n$. The value of $N$ can be estimated by the number of KK modes with mass below $m_{\rm{M}p}$,
\begin{equation}
    N\sim\frac{m_{{\rm M}p}}{m_{\rm KK}}\sim V_{\mathcal{H}}^{\frac{p+1}{7}}\,.
\end{equation}
Considering all KK modes to be uniformly excited up to the species scale, $|f_n|\lesssim M_{\rm Pl,11}=M_{\rm Pl, 4}V_{\mathcal{H}}^{-1/2}$, we then find
\begin{equation}
   v_p^{\frac{4}{(p-7)(p-9)}}V_{\mathcal{H}}^{\frac{2}{p-9}}\sum_{n=1}^N\left(f_n\kappa_4\right)^2\lesssim v_p^{\frac{4}{(p-7)(p-9)}} V_{\mathcal{H}}^{\frac{p+1}{7}+\frac{2}{p-9}-1}=\left\{\begin{array}{ll}
        v_2^{4/35}V_\mathcal{H}^{-6/7}&\text{for }p=2  \\
       v_5^{1/2}V_{\mathcal{H}}^{-9/14} &\text{for }p=5 
   \end{array}\right.\;,
\end{equation}
which by the assumptions in $v_p$ and $V_{\mathcal{H}}$ is expected to be very small.

  With this in mind, we arrive at the following 4-dimensional effective scalar potential
\begin{equation}
    V_{\rm eff}(\rho, t) = M_{\rm Pl,\,4}^4\frac{R_{g_7,0}}{\mathcal{V}_{\mathcal{H},0}^{2/7}}e^{-3\sqrt{\frac{2}{7}}\rho}+\frac{1}{2}\left[m_{\rm M2,\,0}^2 e^{-\sqrt{\frac{2}{7}}\rho}\psi(t)^2+m_{\rm M5,\,0}^2e^{\sqrt{\frac{2}{7}}\rho}\chi(t)^2\right].
\end{equation}
 Since the M5 term grows with $\rho$, our heavy particle will be the M5 brane. The remaining terms are either runaway terms in analogy to section \ref{s.potential and heavy} or light mass terms as in section \ref{s.light and heavy}. The task here is to test whether we can construct transient de Sitter cosmologies with both types of terms.  $V_R$ cannot be made comparable with the M5 mass in a controlled regime, so our only option is the M2-brane tower.\footnote{One could also thread $\mathcal{H}_7$ with an arbitrary number $n_7$ of units of $G_7$ flux. This sources an effective cosmological constant in the $\mathcal{H}_7$ geometry, which after the change to 4d Einstein frame translates into a runaway potential:
$$
    \frac{V_{G_7}(\rho)}{M_{\rm Pl,\,4}^4}=\frac{n_7^2}{2}\exp\left\{-\frac{2\sqrt{14}}{3}\rho\right\}\;.
$$
In a similar manner as with the curvature potential, $V_{G_7}$ cannot be comparable with the M5 term whilst in the controlled regime.}

Under these assumptions, and using the discussion from Section \ref{s.light and heavy}, we arrive at a quasi-dS phase of at most $N=\sqrt{\frac{2}{3}}\simeq 0.82$ $e$-folds, achieved when $\chi$ and $\psi$ become Planckian. We emphasize that this result comes from the free field approximation alone, and that we have neglected both coupling to KK tower as well as quartic effects of the type $\frac{\lambda}{4!}\psi^4$, $\frac{\lambda}{4!}\chi^4$ or $\frac{\lambda}{4!}\psi^2\chi^2$ etc. We will first use dimensional analysis to argue that after compactification to 4d the later mixed terms are suppressed and thus do not pose a threat to the mechanism.

Let $\eta$ be either $\psi$ or $\chi$ (both work the same way), and consider the following term in the 4d EFT action:
\begin{equation}
    S^{(\rm 4d)}\supset-\int\diff^4 x\sqrt{-g}\left(\frac{m^2}{2}\eta^2+\frac{\lambda}{4!}\eta^4\right)\;.
\end{equation}
The quartic term gives rise to an effective potential between two $\eta$ particles. We will now estimate this via the Born approximation in non-relativistic quantum mechanics \cite{TongQFT}. This approximation states that the effect of the quartic interaction can be captured in the non-relativistic regime by an effective potential between two $\eta$ particles of the form 
\begin{equation}U(\vec r)=\frac{\lambda}{4 m^2}\delta^{(3)}(\vec r),\end{equation}
with $\Vec{r}$ the separation between the $\eta$ particles. Taking the effective interaction length of the two-particle system to be its de Broglie wavelength, $L\sim m^{-1}$, this leads to an estimate
\begin{equation}
   \Delta E\sim \frac{\lambda}{4 m^2}m^3\sim \frac{\lambda}{4} m
\end{equation}
for the interaction energy due to the quartic term in the system of two particles. We can compare this with an estimate for $\Delta E$ from the 11-dimensional perspective. Since the particles are wrapped M2 and M5 over different cycles, their intersection is pointlike. This has an interaction energy of order $\Delta E\sim M_{\rm Pl,\,11}$ (since there is no other scale involved in the problem). 
 This results in the identification
\begin{equation}
    \frac{\lambda}{4 m^2}m^3\sim \Delta E\lesssim M_{\rm Pl,\,11}\Longrightarrow \lambda\lesssim\frac{M_{\rm Pl,\,11}}{m}\approx
    v_2^{-1}(R_{\mathcal{H}} M_{\rm Pl,\, 11})^{-2}\approx v_5^{-1}(R_{\mathcal{H}} M_{\rm Pl,\, 11})^{-5}
    \;,
\end{equation}
where we have used \eqref{e.M2M5mass} for the expression of the masses. Note that $\lambda$ is adimensional and for the large volume regime and sub-Planckian vevs of $\chi$ and $\psi$ the quartic terms will be subleading with respect to the mass ones. 

On the other hand, the quartic coupling between two M2's or two M5's is not similarly suppressed. When particles of the same kind coincide in the macroscopic space, they are overlapping also in the internal one, since their associated M$p$-branes are wrapping the same $p$-\emph{torsion} cycle. This can lead to their annihilation, with the interactions generated throughout their shared worldvolume, so we would expect $\Delta E\lesssim M_{11}^{p+1}\upsilon_p R_{\mathcal{H}}^p = m_{{\rm M}p}$, which leads to $\lambda \sim \mathcal{O}(1)$. Given the potential caveats we have considered in this section, these quartic terms can be a real threat to the scenario. As this estimation is based on Born approximation and dimensional analysis, upon more precise calculation it is possible that $\lambda$ actually takes a sufficiently small value.

A term $\lambda(\rho)\eta^4$ would contribute to the effective potential while the massive field $\eta$ is frozen, resulting in Planckian values for the vacuum energy, as explained in Section \ref{s: validity EFT}, for $|\eta|\sim M_{\rm Pl,4}$.
Furthermore, since $\lambda\gg \left(\frac{m_{{\rm M}p}}{M_{\rm Pl,4}}\right)^2$, then the zeroth order term in the $\eta$ e.o.m is dominated by $\lambda\eta^3$ (rather than the mass term $m^2\eta$), greatly reducing the duration of the frozen period and the total number of $e$-folds. Further analysis is beyond the scope of this paper, but the challenges we described appear difficult to overcome.

\section{Conclusions}\label{s: conclusions}
%Please give us the nobel

The difficulty to obtain a cosmological constant in String Theory, some Swampland Conjectures, and now the recent results of the DESI collaboration all serve to motivate the search for alternative cosmological scenarios. In particular, scenarios where the dark energy changes in time and/or where the phase of accelerated expansion just lasts a finite time are especially appealing. In this paper we have focused on one such class of models, that we dubbed ``Cosmological Chameleons'' in analogy with the more standard Chameleon mechanism, where a rolling potential or a light tower is stabilized by the effects of species that becomes heavier as the quintessence field evolves. We explored their viability and stringy embedding. 

We have found that these models can indeed produce transient phases of accelerated expansion, lasting up to $\mathcal{O}(1)$ $e$-folds for high enough densities of the states.  Models of transient acceleration such as these also solve automatically the ``why now'' problem \cite{Agrawal:2018own}, since the phase of accelerated expansion ends in an asymptotic regime where extra dimensions decompactify, ending life as we know it. 

A particularly appealing feature of the scenario is that it requires no novel ingredients, and is built upon standard features present in asymptotic regions of moduli space, where the traditional scalar potential is too steep to accommodate accelerated expansion. The Cosmological Chameleon scenario evades this problem, since the heavy states can stabilize the scalar potential, albeit temporarily.

On the other hand, all the scenarios that we explored for a stringy embedding of the Cosmological Chameleons face significant obstacles. The difficulties are not so different from those that plague dS embeddings and models of quintessence in String Theory, and amount to the fact that it is hard to generate a flat region in a potential where all ingredients are steep, plus fifth force constraints.  Furthermore, the EFT of the heavy states is harder to control, particularly for the Planckian field ranges required to produce $\mathcal{O}(1)$ $e$-folds. Even though interactions are naively expected to be small close to asymptotic regions of moduli space, in some cases they are important enough to jeopardize our construction. One also needs to worry about the coupling to other, light KK towers, to which a small but significant amount of energy may be dumped.

These pose significant challenges to a full realization of the scenario via towers of states, which we did not achieve. Alternatively, there may be other realizations of the Cosmological Chameleon, perhaps involving heavy fields of a different nature to the towers considered here. For instance, one could replace the heavy towers with a monodromic axion \cite{Silverstein:2008sg,McAllister:2008hb}, where interactions for the heavy field can be suppressed to a large extent by the Dvali-Kaloper-Sorbo mechanism \cite{Dvali:2004tma,Dvali:2005an,Kaloper:2008fb,Kaloper:2008qs}. These or other possibilities may turn out to be better stringy embeddings of the Cosmological Chameleon than the ones we considered. In any case, the message is to think outside the box! We do not need to restrict to traditional flux and higher-derivative potential terms, or branes and orientifolds, or Calabi-Yau spaces. Perhaps an accelerating phase in String Theory will be constructed using more exotic ingredients such as hyperbolic manifolds, or towers of heavy fields, or quantum effects. The truth is out there!

\bigskip
{\bf Acknowledgments:}
We would like to thank Yashar Akrami, David Alonso-Gonz\'alez, Alberto Castellano, Damian van de Heisteeg, Bruno de Luca, Joan Quirant, Cumrun Vafa, Bruno Valeixo Bento and Irene Valenzuela  for illuminating discussions. M.M. thanks the KITP Program ``What is String Theory?'' for providing a stimulating enviroment for discussion. This research was supported in part by grant NSF PHY-1748958 to the Kavli Institute for Theoretical Physics (KITP). M.M. and I.R. wish to acknowledge the hospitality of the Department of Theoretical Physics at CERN and the Department of Physics of Harvard University during the different stages of this work. The authors acknowledge the support of the Spanish Agencia Estatal de Investigaci\'on through the grant ``IFT Centro de Excelencia Severo Ochoa'' CEX2020-001007-S and the grant PID2021-123017NB-I00, funded by MCIN/AEI/10.13039/ 501100011033 and by ERDF A way of making Europe. G.F.C. is supported by the grant PRE2021-097279 funded by MCIN/AEI/ 10.13039/501100011033 and by ESF+. MM is supported by an Atraccion del Talento Fellowship 2022-T1/TIC-23956 from Comunidad de Madrid. The work of I.R. is supported by the Spanish FPI grant No. PRE2020-094163.

\vspace*{.5cm}
\bibliographystyle{JHEP2015}
\bibliography{ref}

\providecommand{\href}[2]{#2}\begingroup\raggedright\begin{thebibliography}{100}

\bibitem{Flauger:2022hie}
R.~Flauger, V.~Gorbenko, A.~Joyce, L.~McAllister, G.~Shiu and E.~Silverstein, \emph{{Snowmass White Paper: Cosmology at the Theory Frontier}},  in \emph{{Snowmass 2021}}, 3, 2022, \href{https://arxiv.org/abs/2203.07629}{{\ttfamily 2203.07629}}.

\bibitem{Cicoli:2023opf}
M.~Cicoli, J.~P. Conlon, A.~Maharana, S.~Parameswaran, F.~Quevedo and I.~Zavala, \emph{{String cosmology: From the early universe to today}}, \href{https://doi.org/10.1016/j.physrep.2024.01.002}{\emph{Phys. Rept.} {\bfseries 1059} (2024) 1} [\href{https://arxiv.org/abs/2303.04819}{{\ttfamily 2303.04819}}].

\bibitem{Balasubramanian:2005zx}
V.~Balasubramanian, P.~Berglund, J.~P. Conlon and F.~Quevedo, \emph{{Systematics of moduli stabilisation in Calabi-Yau flux compactifications}}, \href{https://doi.org/10.1088/1126-6708/2005/03/007}{\emph{JHEP} {\bfseries 03} (2005) 007} [\href{https://arxiv.org/abs/hep-th/0502058}{{\ttfamily hep-th/0502058}}].

\bibitem{deAlwis:2014wia}
S.~de~Alwis, R.~K. Gupta, F.~Quevedo and R.~Valandro, \emph{{On KKLT/CFT and LVS/CFT Dualities}}, \href{https://doi.org/10.1007/JHEP07(2015)036}{\emph{JHEP} {\bfseries 07} (2015) 036} [\href{https://arxiv.org/abs/1412.6999}{{\ttfamily 1412.6999}}].

\bibitem{Polchinski:2015bea}
J.~Polchinski, \emph{{Brane/antibrane dynamics and KKLT stability}},  \href{https://arxiv.org/abs/1509.05710}{{\ttfamily 1509.05710}}.

\bibitem{Sethi:2017phn}
S.~Sethi, \emph{{Supersymmetry Breaking by Fluxes}}, \href{https://doi.org/10.1007/JHEP10(2018)022}{\emph{JHEP} {\bfseries 10} (2018) 022} [\href{https://arxiv.org/abs/1709.03554}{{\ttfamily 1709.03554}}].

\bibitem{Gao:2020xqh}
X.~Gao, A.~Hebecker and D.~Junghans, \emph{{Control issues of KKLT}}, \href{https://doi.org/10.1002/prop.202000089}{\emph{Fortsch. Phys.} {\bfseries 68} (2020) 2000089} [\href{https://arxiv.org/abs/2009.03914}{{\ttfamily 2009.03914}}].

\bibitem{Bena:2020xrh}
I.~Bena, J.~Bl\r{a}b\"ack, M.~Gra\~na and S.~L\"ust, \emph{{The tadpole problem}}, \href{https://doi.org/10.1007/JHEP11(2021)223}{\emph{JHEP} {\bfseries 11} (2021) 223} [\href{https://arxiv.org/abs/2010.10519}{{\ttfamily 2010.10519}}].

\bibitem{Lust:2022lfc}
S.~L\"ust, C.~Vafa, M.~Wiesner and K.~Xu, \emph{{Holography and the KKLT scenario}}, \href{https://doi.org/10.1007/JHEP10(2022)188}{\emph{JHEP} {\bfseries 10} (2022) 188} [\href{https://arxiv.org/abs/2204.07171}{{\ttfamily 2204.07171}}].

\bibitem{Grana:2022nyp}
M.~Gra\~na, N.~Kovensky and D.~Toulikas, \emph{{Smearing and unsmearing KKLT AdS vacua}}, \href{https://doi.org/10.1007/JHEP03(2023)015}{\emph{JHEP} {\bfseries 03} (2023) 015} [\href{https://arxiv.org/abs/2212.05074}{{\ttfamily 2212.05074}}].

\bibitem{Blumenhagen:2022dbo}
R.~Blumenhagen, A.~Gligovic and S.~Kaddachi, \emph{{Mass Hierarchies and Quantum Gravity Constraints in DKMM-refined KKLT}}, \href{https://doi.org/10.1002/prop.202200167}{\emph{Fortsch. Phys.} {\bfseries 71} (2023) 2200167} [\href{https://arxiv.org/abs/2206.08400}{{\ttfamily 2206.08400}}].

\bibitem{Lust:2022xoq}
S.~L\"ust and L.~Randall, \emph{{Effective Theory of Warped Compactifications and the Implications for KKLT}}, \href{https://doi.org/10.1002/prop.202200103}{\emph{Fortsch. Phys.} {\bfseries 70} (2022) 2200103} [\href{https://arxiv.org/abs/2206.04708}{{\ttfamily 2206.04708}}].

\bibitem{Kachru:2003aw}
S.~Kachru, R.~Kallosh, A.~D. Linde and S.~P. Trivedi, \emph{{De Sitter vacua in string theory}}, \href{https://doi.org/10.1103/PhysRevD.68.046005}{\emph{Phys. Rev. D} {\bfseries 68} (2003) 046005} [\href{https://arxiv.org/abs/hep-th/0301240}{{\ttfamily hep-th/0301240}}].

\bibitem{Vafa:2005ui}
C.~Vafa, \emph{{The String landscape and the swampland}},  \href{https://arxiv.org/abs/hep-th/0509212}{{\ttfamily hep-th/0509212}}.

\bibitem{Brennan:2017rbf}
T.~D. Brennan, F.~Carta and C.~Vafa, \emph{{The String Landscape, the Swampland, and the Missing Corner}}, \href{https://doi.org/10.22323/1.305.0015}{\emph{PoS} {\bfseries TASI2017} (2017) 015} [\href{https://arxiv.org/abs/1711.00864}{{\ttfamily 1711.00864}}].

\bibitem{Palti:2019pca}
E.~Palti, \emph{{The Swampland: Introduction and Review}}, \href{https://doi.org/10.1002/prop.201900037}{\emph{Fortsch. Phys.} {\bfseries 67} (2019) 1900037} [\href{https://arxiv.org/abs/1903.06239}{{\ttfamily 1903.06239}}].

\bibitem{vanBeest:2021lhn}
M.~van Beest, J.~Calder\'on-Infante, D.~Mirfendereski and I.~Valenzuela, \emph{{Lectures on the Swampland Program in String Compactifications}}, \href{https://doi.org/10.1016/j.physrep.2022.09.002}{\emph{Phys. Rept.} {\bfseries 989} (2022) 1} [\href{https://arxiv.org/abs/2102.01111}{{\ttfamily 2102.01111}}].

\bibitem{Grana:2021zvf}
M.~Gra\~na and A.~Herr\'aez, \emph{{The Swampland Conjectures: A Bridge from Quantum Gravity to Particle Physics}}, \href{https://doi.org/10.3390/universe7080273}{\emph{Universe} {\bfseries 7} (2021) 273} [\href{https://arxiv.org/abs/2107.00087}{{\ttfamily 2107.00087}}].

\bibitem{Harlow:2022ich}
D.~Harlow, B.~Heidenreich, M.~Reece and T.~Rudelius, \emph{{Weak gravity conjecture}}, \href{https://doi.org/10.1103/RevModPhys.95.035003}{\emph{Rev. Mod. Phys.} {\bfseries 95} (2023) 035003} [\href{https://arxiv.org/abs/2201.08380}{{\ttfamily 2201.08380}}].

\bibitem{Agmon:2022thq}
N.~B. Agmon, A.~Bedroya, M.~J. Kang and C.~Vafa, \emph{{Lectures on the string landscape and the Swampland}},  \href{https://arxiv.org/abs/2212.06187}{{\ttfamily 2212.06187}}.

\bibitem{VanRiet:2023pnx}
T.~Van~Riet and G.~Zoccarato, \emph{{Beginners lectures on flux compactifications and related Swampland topics}}, \href{https://doi.org/10.1016/j.physrep.2023.11.003}{\emph{Phys. Rept.} {\bfseries 1049} (2024) 1} [\href{https://arxiv.org/abs/2305.01722}{{\ttfamily 2305.01722}}].

\bibitem{Obied:2018sgi}
G.~Obied, H.~Ooguri, L.~Spodyneiko and C.~Vafa, \emph{{De Sitter Space and the Swampland}},  \href{https://arxiv.org/abs/1806.08362}{{\ttfamily 1806.08362}}.

\bibitem{DESI:2024mwx}
{\scshape DESI} collaboration, A.~G. Adame et~al., \emph{{DESI 2024 VI: Cosmological Constraints from the Measurements of Baryon Acoustic Oscillations}},  \href{https://arxiv.org/abs/2404.03002}{{\ttfamily 2404.03002}}.

\bibitem{Tsujikawa:2013fta}
S.~Tsujikawa, \emph{{Quintessence: A Review}}, \href{https://doi.org/10.1088/0264-9381/30/21/214003}{\emph{Class. Quant. Grav.} {\bfseries 30} (2013) 214003} [\href{https://arxiv.org/abs/1304.1961}{{\ttfamily 1304.1961}}].

\bibitem{Achucarro:2018vey}
A.~Ach\'ucarro and G.~A. Palma, \emph{{The string swampland constraints require multi-field inflation}}, \href{https://doi.org/10.1088/1475-7516/2019/02/041}{\emph{JCAP} {\bfseries 02} (2019) 041} [\href{https://arxiv.org/abs/1807.04390}{{\ttfamily 1807.04390}}].

\bibitem{Planck:2018vyg}
{\scshape Planck} collaboration, N.~Aghanim et~al., \emph{{Planck 2018 results. VI. Cosmological parameters}}, \href{https://doi.org/10.1051/0004-6361/201833910}{\emph{Astron. Astrophys.} {\bfseries 641} (2020) A6} [\href{https://arxiv.org/abs/1807.06209}{{\ttfamily 1807.06209}}].

\bibitem{Gomes:2023dat}
J.~M. Gomes, E.~Hardy and S.~Parameswaran, \emph{{Dark Energy with a Little Help from its Friends}},  \href{https://arxiv.org/abs/2311.08888}{{\ttfamily 2311.08888}}.

\bibitem{Khoury:2003rn}
J.~Khoury and A.~Weltman, \emph{{Chameleon cosmology}}, \href{https://doi.org/10.1103/PhysRevD.69.044026}{\emph{Phys. Rev. D} {\bfseries 69} (2004) 044026} [\href{https://arxiv.org/abs/astro-ph/0309411}{{\ttfamily astro-ph/0309411}}].

\bibitem{Khoury:2003aq}
J.~Khoury and A.~Weltman, \emph{{Chameleon fields: Awaiting surprises for tests of gravity in space}}, \href{https://doi.org/10.1103/PhysRevLett.93.171104}{\emph{Phys. Rev. Lett.} {\bfseries 93} (2004) 171104} [\href{https://arxiv.org/abs/astro-ph/0309300}{{\ttfamily astro-ph/0309300}}].

\bibitem{Ooguri:2006in}
H.~Ooguri and C.~Vafa, \emph{{On the Geometry of the String Landscape and the Swampland}}, \href{https://doi.org/10.1016/j.nuclphysb.2006.10.033}{\emph{Nucl.Phys.} {\bfseries B766} (2007) 21} [\href{https://arxiv.org/abs/hep-th/0605264}{{\ttfamily hep-th/0605264}}].

\bibitem{Baume:2016psm}
F.~Baume and E.~Palti, \emph{{Backreacted Axion Field Ranges in String Theory}}, \href{https://doi.org/10.1007/JHEP08(2016)043}{\emph{JHEP} {\bfseries 08} (2016) 043} [\href{https://arxiv.org/abs/1602.06517}{{\ttfamily 1602.06517}}].

\bibitem{Klaewer:2016kiy}
D.~Klaewer and E.~Palti, \emph{{Super-Planckian Spatial Field Variations and Quantum Gravity}}, \href{https://doi.org/10.1007/JHEP01(2017)088}{\emph{JHEP} {\bfseries 01} (2017) 088} [\href{https://arxiv.org/abs/1610.00010}{{\ttfamily 1610.00010}}].

\bibitem{Blumenhagen:2017cxt}
R.~Blumenhagen, I.~Valenzuela and F.~Wolf, \emph{{The Swampland Conjecture and F-term Axion Monodromy Inflation}}, \href{https://doi.org/10.1007/JHEP07(2017)145}{\emph{JHEP} {\bfseries 07} (2017) 145} [\href{https://arxiv.org/abs/1703.05776}{{\ttfamily 1703.05776}}].

\bibitem{Grimm:2018ohb}
T.~W. Grimm, E.~Palti and I.~Valenzuela, \emph{{Infinite Distances in Field Space and Massless Towers of States}}, \href{https://doi.org/10.1007/JHEP08(2018)143}{\emph{JHEP} {\bfseries 08} (2018) 143} [\href{https://arxiv.org/abs/1802.08264}{{\ttfamily 1802.08264}}].

\bibitem{Heidenreich:2018kpg}
B.~Heidenreich, M.~Reece and T.~Rudelius, \emph{{Emergence of Weak Coupling at Large Distance in Quantum Gravity}}, \href{https://doi.org/10.1103/PhysRevLett.121.051601}{\emph{Phys. Rev. Lett.} {\bfseries 121} (2018) 051601} [\href{https://arxiv.org/abs/1802.08698}{{\ttfamily 1802.08698}}].

\bibitem{Blumenhagen:2018nts}
R.~Blumenhagen, D.~Kl\"awer, L.~Schlechter and F.~Wolf, \emph{{The Refined Swampland Distance Conjecture in Calabi-Yau Moduli Spaces}}, \href{https://doi.org/10.1007/JHEP06(2018)052}{\emph{JHEP} {\bfseries 06} (2018) 052} [\href{https://arxiv.org/abs/1803.04989}{{\ttfamily 1803.04989}}].

\bibitem{Grimm:2018cpv}
T.~W. Grimm, C.~Li and E.~Palti, \emph{{Infinite Distance Networks in Field Space and Charge Orbits}}, \href{https://doi.org/10.1007/JHEP03(2019)016}{\emph{JHEP} {\bfseries 03} (2019) 016} [\href{https://arxiv.org/abs/1811.02571}{{\ttfamily 1811.02571}}].

\bibitem{Buratti:2018xjt}
G.~Buratti, J.~Calder\'on and A.~M. Uranga, \emph{{Transplanckian axion monodromy!?}}, \href{https://doi.org/10.1007/JHEP05(2019)176}{\emph{JHEP} {\bfseries 05} (2019) 176} [\href{https://arxiv.org/abs/1812.05016}{{\ttfamily 1812.05016}}].

\bibitem{Corvilain:2018lgw}
P.~Corvilain, T.~W. Grimm and I.~Valenzuela, \emph{{The Swampland Distance Conjecture for K\"ahler moduli}}, \href{https://doi.org/10.1007/JHEP08(2019)075}{\emph{JHEP} {\bfseries 08} (2019) 075} [\href{https://arxiv.org/abs/1812.07548}{{\ttfamily 1812.07548}}].

\bibitem{Joshi:2019nzi}
A.~Joshi and A.~Klemm, \emph{{Swampland Distance Conjecture for One-Parameter Calabi-Yau Threefolds}}, \href{https://doi.org/10.1007/JHEP08(2019)086}{\emph{JHEP} {\bfseries 08} (2019) 086} [\href{https://arxiv.org/abs/1903.00596}{{\ttfamily 1903.00596}}].

\bibitem{Erkinger:2019umg}
D.~Erkinger and J.~Knapp, \emph{{Refined swampland distance conjecture and exotic hybrid Calabi-Yaus}}, \href{https://doi.org/10.1007/JHEP07(2019)029}{\emph{JHEP} {\bfseries 07} (2019) 029} [\href{https://arxiv.org/abs/1905.05225}{{\ttfamily 1905.05225}}].

\bibitem{Marchesano:2019ifh}
F.~Marchesano and M.~Wiesner, \emph{{Instantons and infinite distances}}, \href{https://doi.org/10.1007/JHEP08(2019)088}{\emph{JHEP} {\bfseries 08} (2019) 088} [\href{https://arxiv.org/abs/1904.04848}{{\ttfamily 1904.04848}}].

\bibitem{Font:2019cxq}
A.~Font, A.~Herr\'aez and L.~E. Ib\'a\~nez, \emph{{The Swampland Distance Conjecture and Towers of Tensionless Branes}}, \href{https://doi.org/10.1007/JHEP08(2019)044}{\emph{JHEP} {\bfseries 08} (2019) 044} [\href{https://arxiv.org/abs/1904.05379}{{\ttfamily 1904.05379}}].

\bibitem{Gendler:2020dfp}
N.~Gendler and I.~Valenzuela, \emph{{Merging the weak gravity and distance conjectures using BPS extremal black holes}}, \href{https://doi.org/10.1007/JHEP01(2021)176}{\emph{JHEP} {\bfseries 01} (2021) 176} [\href{https://arxiv.org/abs/2004.10768}{{\ttfamily 2004.10768}}].

\bibitem{Lanza:2020qmt}
S.~Lanza, F.~Marchesano, L.~Martucci and I.~Valenzuela, \emph{{Swampland Conjectures for Strings and Membranes}}, \href{https://doi.org/10.1007/JHEP02(2021)006}{\emph{JHEP} {\bfseries 02} (2021) 006} [\href{https://arxiv.org/abs/2006.15154}{{\ttfamily 2006.15154}}].

\bibitem{Klaewer:2020lfg}
D.~Klaewer, S.-J. Lee, T.~Weigand and M.~Wiesner, \emph{{Quantum Corrections in 4d N=1 Infinite Distance Limits and the Weak Gravity Conjecture}},  \href{https://arxiv.org/abs/2011.00024}{{\ttfamily 2011.00024}}.

\bibitem{Rudelius:2023mjy}
T.~Rudelius, \emph{{Revisiting the Refined Distance Conjecture}},  \href{https://arxiv.org/abs/2303.12103}{{\ttfamily 2303.12103}}.

\bibitem{Calderon-Infante:2023ler}
J.~Calder\'on-Infante, A.~Castellano, A.~Herr\'aez and L.~E. Ib\'a\~nez, \emph{{Entropy bounds and the species scale distance conjecture}}, \href{https://doi.org/10.1007/JHEP01(2024)039}{\emph{JHEP} {\bfseries 01} (2024) 039} [\href{https://arxiv.org/abs/2306.16450}{{\ttfamily 2306.16450}}].

\bibitem{Ooguri:2024ofs}
H.~Ooguri and Y.~Wang, \emph{{Universal Bounds on CFT Distance Conjecture}},  \href{https://arxiv.org/abs/2405.00674}{{\ttfamily 2405.00674}}.

\bibitem{Aoufia:2024awo}
C.~Aoufia, I.~Basile and G.~Leone, \emph{{Species scale, worldsheet CFTs and emergent geometry}},  \href{https://arxiv.org/abs/2405.03683}{{\ttfamily 2405.03683}}.

\bibitem{Lee:2019wij}
S.-J. Lee, W.~Lerche and T.~Weigand, \emph{{Emergent Strings from Infinite Distance Limits}},  \href{https://arxiv.org/abs/1910.01135}{{\ttfamily 1910.01135}}.

\bibitem{Rudelius:2021azq}
T.~Rudelius, \emph{{Asymptotic observables and the swampland}}, \href{https://doi.org/10.1103/PhysRevD.104.126023}{\emph{Phys. Rev. D} {\bfseries 104} (2021) 126023} [\href{https://arxiv.org/abs/2106.09026}{{\ttfamily 2106.09026}}].

\bibitem{Calderon-Infante:2020dhm}
J.~Calder\'on-Infante, A.~M. Uranga and I.~Valenzuela, \emph{{The Convex Hull Swampland Distance Conjecture and Bounds on Non-geodesics}}, \href{https://doi.org/10.1007/JHEP03(2021)299}{\emph{JHEP} {\bfseries 03} (2021) 299} [\href{https://arxiv.org/abs/2012.00034}{{\ttfamily 2012.00034}}].

\bibitem{Grimm:2022sbl}
T.~W. Grimm, S.~Lanza and C.~Li, \emph{{Tameness, Strings, and the Distance Conjecture}}, \href{https://doi.org/10.1007/JHEP09(2022)149}{\emph{JHEP} {\bfseries 09} (2022) 149} [\href{https://arxiv.org/abs/2206.00697}{{\ttfamily 2206.00697}}].

\bibitem{Calderon-Infante:2022nxb}
J.~Calder\'on-Infante, I.~Ruiz and I.~Valenzuela, \emph{{Asymptotic accelerated expansion in string theory and the Swampland}}, \href{https://doi.org/10.1007/JHEP06(2023)129}{\emph{JHEP} {\bfseries 06} (2023) 129} [\href{https://arxiv.org/abs/2209.11821}{{\ttfamily 2209.11821}}].

\bibitem{Maldacena:2000mw}
J.~M. Maldacena and C.~Nunez, \emph{{Supergravity description of field theories on curved manifolds and a no go theorem}}, \href{https://doi.org/10.1142/S0217751X01003937}{\emph{Int. J. Mod. Phys. A} {\bfseries 16} (2001) 822} [\href{https://arxiv.org/abs/hep-th/0007018}{{\ttfamily hep-th/0007018}}].

\bibitem{Hertzberg:2007wc}
M.~P. Hertzberg, S.~Kachru, W.~Taylor and M.~Tegmark, \emph{{Inflationary Constraints on Type IIA String Theory}}, \href{https://doi.org/10.1088/1126-6708/2007/12/095}{\emph{JHEP} {\bfseries 12} (2007) 095} [\href{https://arxiv.org/abs/0711.2512}{{\ttfamily 0711.2512}}].

\bibitem{Andriot:2019wrs}
D.~Andriot, \emph{{Open problems on classical de Sitter solutions}}, \href{https://doi.org/10.1002/prop.201900026}{\emph{Fortsch. Phys.} {\bfseries 67} (2019) 1900026} [\href{https://arxiv.org/abs/1902.10093}{{\ttfamily 1902.10093}}].

\bibitem{Andriot:2020lea}
D.~Andriot, N.~Cribiori and D.~Erkinger, \emph{{The web of swampland conjectures and the TCC bound}}, \href{https://doi.org/10.1007/JHEP07(2020)162}{\emph{JHEP} {\bfseries 07} (2020) 162} [\href{https://arxiv.org/abs/2004.00030}{{\ttfamily 2004.00030}}].

\bibitem{Shiu:2023fhb}
G.~Shiu, F.~Tonioni and H.~V. Tran, \emph{{Late-time attractors and cosmic acceleration}}, \href{https://doi.org/10.1103/PhysRevD.108.063528}{\emph{Phys. Rev. D} {\bfseries 108} (2023) 063528} [\href{https://arxiv.org/abs/2306.07327}{{\ttfamily 2306.07327}}].

\bibitem{Shiu:2023nph}
G.~Shiu, F.~Tonioni and H.~V. Tran, \emph{{Accelerating universe at the end of time}}, \href{https://doi.org/10.1103/PhysRevD.108.063527}{\emph{Phys. Rev. D} {\bfseries 108} (2023) 063527} [\href{https://arxiv.org/abs/2303.03418}{{\ttfamily 2303.03418}}].

\bibitem{Cremonini:2023suw}
S.~Cremonini, E.~Gonzalo, M.~Rajaguru, Y.~Tang and T.~Wrase, \emph{{On asymptotic dark energy in string theory}}, \href{https://doi.org/10.1007/JHEP09(2023)075}{\emph{JHEP} {\bfseries 09} (2023) 075} [\href{https://arxiv.org/abs/2306.15714}{{\ttfamily 2306.15714}}].

\bibitem{Hebecker:2023qke}
A.~Hebecker, S.~Schreyer and G.~Venken, \emph{{No asymptotic acceleration without higher-dimensional de Sitter vacua}}, \href{https://doi.org/10.1007/JHEP11(2023)173}{\emph{JHEP} {\bfseries 11} (2023) 173} [\href{https://arxiv.org/abs/2306.17213}{{\ttfamily 2306.17213}}].

\bibitem{VanRiet:2023cca}
T.~Van~Riet, \emph{{No accelerating scaling cosmologies at string tree level?}}, \href{https://doi.org/10.1088/1475-7516/2024/01/055}{\emph{JCAP} {\bfseries 01} (2024) 055} [\href{https://arxiv.org/abs/2308.15035}{{\ttfamily 2308.15035}}].

\bibitem{Seo:2024fki}
M.-S. Seo, \emph{{Asymptotic bound on slow-roll parameter in stringy quintessence model}},  \href{https://arxiv.org/abs/2402.00241}{{\ttfamily 2402.00241}}.

\bibitem{Bedroya:2019snp}
A.~Bedroya and C.~Vafa, \emph{{Trans-Planckian Censorship and the Swampland}}, \href{https://doi.org/10.1007/JHEP09(2020)123}{\emph{JHEP} {\bfseries 09} (2020) 123} [\href{https://arxiv.org/abs/1909.11063}{{\ttfamily 1909.11063}}].

\bibitem{Etheredge:2023odp}
M.~Etheredge, B.~Heidenreich, J.~McNamara, T.~Rudelius, I.~Ruiz and I.~Valenzuela, \emph{{Running decompactification, sliding towers, and the distance conjecture}}, \href{https://doi.org/10.1007/JHEP12(2023)182}{\emph{JHEP} {\bfseries 12} (2023) 182} [\href{https://arxiv.org/abs/2306.16440}{{\ttfamily 2306.16440}}].

\bibitem{Etheredge:2023usk}
M.~Etheredge, \emph{{Dense geodesics, tower alignment, and the Sharpened Distance Conjecture}}, \href{https://doi.org/10.1007/JHEP01(2024)122}{\emph{JHEP} {\bfseries 01} (2024) 122} [\href{https://arxiv.org/abs/2308.01331}{{\ttfamily 2308.01331}}].

\bibitem{Adelberger:2003zx}
E.~G. Adelberger, B.~R. Heckel and A.~E. Nelson, \emph{{Tests of the gravitational inverse square law}}, \href{https://doi.org/10.1146/annurev.nucl.53.041002.110503}{\emph{Ann. Rev. Nucl. Part. Sci.} {\bfseries 53} (2003) 77} [\href{https://arxiv.org/abs/hep-ph/0307284}{{\ttfamily hep-ph/0307284}}].

\bibitem{Etheredge:2022opl}
M.~Etheredge, B.~Heidenreich, S.~Kaya, Y.~Qiu and T.~Rudelius, \emph{{Sharpening the Distance Conjecture in Diverse Dimensions}},  \href{https://arxiv.org/abs/2206.04063}{{\ttfamily 2206.04063}}.

\bibitem{Bachlechner:2014gfa}
T.~C. Bachlechner, C.~Long and L.~McAllister, \emph{{Planckian Axions in String Theory}}, \href{https://doi.org/10.1007/JHEP12(2015)042}{\emph{JHEP} {\bfseries 12} (2015) 042} [\href{https://arxiv.org/abs/1412.1093}{{\ttfamily 1412.1093}}].

\bibitem{Bachlechner:2015qja}
T.~C. Bachlechner, C.~Long and L.~McAllister, \emph{{Planckian Axions and the Weak Gravity Conjecture}}, \href{https://doi.org/10.1007/JHEP01(2016)091}{\emph{JHEP} {\bfseries 01} (2016) 091} [\href{https://arxiv.org/abs/1503.07853}{{\ttfamily 1503.07853}}].

\bibitem{Hebecker:2015rya}
A.~Hebecker, P.~Mangat, F.~Rompineve and L.~T. Witkowski, \emph{{Winding out of the Swamp: Evading the Weak Gravity Conjecture with F-term Winding Inflation?}}, \href{https://doi.org/10.1016/j.physletb.2015.07.026}{\emph{Phys. Lett. B} {\bfseries 748} (2015) 455} [\href{https://arxiv.org/abs/1503.07912}{{\ttfamily 1503.07912}}].

\bibitem{Montero:2015ofa}
M.~Montero, A.~M. Uranga and I.~Valenzuela, \emph{{Transplanckian axions!?}}, \href{https://doi.org/10.1007/JHEP08(2015)032}{\emph{JHEP} {\bfseries 08} (2015) 032} [\href{https://arxiv.org/abs/1503.03886}{{\ttfamily 1503.03886}}].

\bibitem{PhysRev.21.483}
A.~H. Compton, \emph{A quantum theory of the scattering of x-rays by light elements}, \href{https://doi.org/10.1103/PhysRev.21.483}{\emph{Phys. Rev.} {\bfseries 21} (1923) 483}.

\bibitem{TongQFT}
D.~Tong, \emph{Quantum field theory},  December, 2007.

\bibitem{Dvali:2007hz}
G.~Dvali, \emph{{Black Holes and Large N Species Solution to the Hierarchy Problem}}, \href{https://doi.org/10.1002/prop.201000009}{\emph{Fortsch. Phys.} {\bfseries 58} (2010) 528} [\href{https://arxiv.org/abs/0706.2050}{{\ttfamily 0706.2050}}].

\bibitem{Dvali:2007wp}
G.~Dvali and M.~Redi, \emph{{Black Hole Bound on the Number of Species and Quantum Gravity at LHC}}, \href{https://doi.org/10.1103/PhysRevD.77.045027}{\emph{Phys. Rev.} {\bfseries D77} (2008) 045027} [\href{https://arxiv.org/abs/0710.4344}{{\ttfamily 0710.4344}}].

\bibitem{Dvali:2008ec}
G.~Dvali and C.~Gomez, \emph{{Quantum Information and Gravity Cutoff in Theories with Species}}, \href{https://doi.org/10.1016/j.physletb.2009.03.024}{\emph{Phys. Lett. B} {\bfseries 674} (2009) 303} [\href{https://arxiv.org/abs/0812.1940}{{\ttfamily 0812.1940}}].

\bibitem{Long:2021jlv}
C.~Long, M.~Montero, C.~Vafa and I.~Valenzuela, \emph{{The desert and the swampland}}, \href{https://doi.org/10.1007/JHEP03(2023)109}{\emph{JHEP} {\bfseries 03} (2023) 109} [\href{https://arxiv.org/abs/2112.11467}{{\ttfamily 2112.11467}}].

\bibitem{vandeHeisteeg:2022btw}
D.~van~de Heisteeg, C.~Vafa, M.~Wiesner and D.~H. Wu, \emph{{Moduli-dependent Species Scale}},  \href{https://arxiv.org/abs/2212.06841}{{\ttfamily 2212.06841}}.

\bibitem{vandeHeisteeg:2023ubh}
D.~van~de Heisteeg, C.~Vafa and M.~Wiesner, \emph{{Bounds on Species Scale and the Distance Conjecture}}, \href{https://doi.org/10.1002/prop.202300143}{\emph{Fortsch. Phys.} {\bfseries 71} (2023) 2300143} [\href{https://arxiv.org/abs/2303.13580}{{\ttfamily 2303.13580}}].

\bibitem{Gkountoumis:2023fym}
G.~Gkountoumis, C.~Hull, K.~Stemerdink and S.~Vandoren, \emph{{Freely acting orbifolds of type IIB string theory on T$^{5}$}}, \href{https://doi.org/10.1007/JHEP08(2023)089}{\emph{JHEP} {\bfseries 08} (2023) 089} [\href{https://arxiv.org/abs/2302.09112}{{\ttfamily 2302.09112}}].

\bibitem{Arkani-Hamed:2007ryu}
N.~Arkani-Hamed, S.~Dubovsky, A.~Nicolis and G.~Villadoro, \emph{{Quantum Horizons of the Standard Model Landscape}}, \href{https://doi.org/10.1088/1126-6708/2007/06/078}{\emph{JHEP} {\bfseries 06} (2007) 078} [\href{https://arxiv.org/abs/hep-th/0703067}{{\ttfamily hep-th/0703067}}].

\bibitem{Grimm:2004ua}
T.~W. Grimm and J.~Louis, \emph{{The Effective action of type IIA Calabi-Yau orientifolds}}, \href{https://doi.org/10.1016/j.nuclphysb.2005.04.007}{\emph{Nucl. Phys. B} {\bfseries 718} (2005) 153} [\href{https://arxiv.org/abs/hep-th/0412277}{{\ttfamily hep-th/0412277}}].

\bibitem{DeWolfe:2005uu}
O.~DeWolfe, A.~Giryavets, S.~Kachru and W.~Taylor, \emph{{Type IIA moduli stabilization}}, \href{https://doi.org/10.1088/1126-6708/2005/07/066}{\emph{JHEP} {\bfseries 07} (2005) 066} [\href{https://arxiv.org/abs/hep-th/0505160}{{\ttfamily hep-th/0505160}}].

\bibitem{Camara:2005dc}
P.~G. Camara, A.~Font and L.~E. Ibanez, \emph{{Fluxes, moduli fixing and MSSM-like vacua in a simple IIA orientifold}}, \href{https://doi.org/10.1088/1126-6708/2005/09/013}{\emph{JHEP} {\bfseries 09} (2005) 013} [\href{https://arxiv.org/abs/hep-th/0506066}{{\ttfamily hep-th/0506066}}].

\bibitem{Polchinski:1998rr}
J.~Polchinski, \emph{{String theory. Vol. 2: Superstring theory and beyond}}, Cambridge Monographs on Mathematical Physics. Cambridge University Press, 12, 2007, \href{https://doi.org/10.1017/CBO9780511618123}{10.1017/CBO9780511618123}.

\bibitem{Strominger:1995cz}
A.~Strominger, \emph{{Massless black holes and conifolds in string theory}}, \href{https://doi.org/10.1016/0550-3213(95)00287-3}{\emph{Nucl. Phys. B} {\bfseries 451} (1995) 96} [\href{https://arxiv.org/abs/hep-th/9504090}{{\ttfamily hep-th/9504090}}].

\bibitem{Townsend:2003fx}
P.~K. Townsend and M.~N.~R. Wohlfarth, \emph{{Accelerating cosmologies from compactification}}, \href{https://doi.org/10.1103/PhysRevLett.91.061302}{\emph{Phys. Rev. Lett.} {\bfseries 91} (2003) 061302} [\href{https://arxiv.org/abs/hep-th/0303097}{{\ttfamily hep-th/0303097}}].

\bibitem{Ohta:2004wk}
N.~Ohta, \emph{{Accelerating cosmologies and inflation from M/superstring theories}}, \href{https://doi.org/10.1142/S0217751X05021257}{\emph{Int. J. Mod. Phys. A} {\bfseries 20} (2005) 1} [\href{https://arxiv.org/abs/hep-th/0411230}{{\ttfamily hep-th/0411230}}].

\bibitem{Castellano:2023jjt}
A.~Castellano, I.~Ruiz and I.~Valenzuela, \emph{{Stringy Evidence for a Universal Pattern at Infinite Distance}},  \href{https://arxiv.org/abs/2311.01536}{{\ttfamily 2311.01536}}.

\bibitem{porti2016reidemeister}
J.~Porti, \emph{Reidemeister torsion, hyperbolic three-manifolds, and character varieties},  2016.

\bibitem{Agrawal:2018own}
P.~Agrawal, G.~Obied, P.~J. Steinhardt and C.~Vafa, \emph{{On the Cosmological Implications of the String Swampland}}, \href{https://doi.org/10.1016/j.physletb.2018.07.040}{\emph{Phys. Lett. B} {\bfseries 784} (2018) 271} [\href{https://arxiv.org/abs/1806.09718}{{\ttfamily 1806.09718}}].

\bibitem{Silverstein:2008sg}
E.~Silverstein and A.~Westphal, \emph{{Monodromy in the CMB: Gravity Waves and String Inflation}}, \href{https://doi.org/10.1103/PhysRevD.78.106003}{\emph{Phys. Rev. D} {\bfseries 78} (2008) 106003} [\href{https://arxiv.org/abs/0803.3085}{{\ttfamily 0803.3085}}].

\bibitem{McAllister:2008hb}
L.~McAllister, E.~Silverstein and A.~Westphal, \emph{{Gravity Waves and Linear Inflation from Axion Monodromy}}, \href{https://doi.org/10.1103/PhysRevD.82.046003}{\emph{Phys. Rev. D} {\bfseries 82} (2010) 046003} [\href{https://arxiv.org/abs/0808.0706}{{\ttfamily 0808.0706}}].

\bibitem{Dvali:2004tma}
G.~Dvali, \emph{{Large hierarchies from attractor vacua}}, \href{https://doi.org/10.1103/PhysRevD.74.025018}{\emph{Phys. Rev. D} {\bfseries 74} (2006) 025018} [\href{https://arxiv.org/abs/hep-th/0410286}{{\ttfamily hep-th/0410286}}].

\bibitem{Dvali:2005an}
G.~Dvali, \emph{{Three-form gauging of axion symmetries and gravity}},  \href{https://arxiv.org/abs/hep-th/0507215}{{\ttfamily hep-th/0507215}}.

\bibitem{Kaloper:2008fb}
N.~Kaloper and L.~Sorbo, \emph{{A Natural Framework for Chaotic Inflation}}, \href{https://doi.org/10.1103/PhysRevLett.102.121301}{\emph{Phys. Rev. Lett.} {\bfseries 102} (2009) 121301} [\href{https://arxiv.org/abs/0811.1989}{{\ttfamily 0811.1989}}].

\bibitem{Kaloper:2008qs}
N.~Kaloper and L.~Sorbo, \emph{{Where in the String Landscape is Quintessence}}, \href{https://doi.org/10.1103/PhysRevD.79.043528}{\emph{Phys. Rev. D} {\bfseries 79} (2009) 043528} [\href{https://arxiv.org/abs/0810.5346}{{\ttfamily 0810.5346}}].

\end{thebibliography}\endgroup

\end{document}